\documentclass[paper, 12pt]{JHEP} \usepackage{amsmath}
\usepackage{amssymb}

\def\Bbb{\mathbb} \def\C{{\Bbb C}} \def\R{{\Bbb R}} 
 \def\Q{{\Bbb Q}} \def\P{{\Bbb P}}

  \def\vol {{\rm
    vol}}  \def\tr{\operatorname{tr}}

 \def\tr{{\rm tr\, }} \def\dim{{\rm dim}}
 \def\Tr{{\rm Tr\, }} 
   \def\deg{{\rm deg}} \def\det{{\rm det}}

\def\id{\protect{{1 \kern-.28em {\rm l}}}}

\newcommand{\be}{\begin{equation}} \newcommand{\ee}{\end{equation}}
\newcommand{\bea}{\begin{eqnarray}} \newcommand{\eea}{\end{eqnarray}}
\newcommand{\beann}{\begin{eqnarray*}}
  \newcommand{\eeann}{\end{eqnarray*}}
\newcommand{\bfig}{\begin{figure}} \newcommand{\efig}{\end{figure}}
\newcommand{\nn}{\nonumber}
\newcommand{\ba}{\begin{array}}\newcommand{\ea}{\end{array}}

\newtheorem{Proposition}{Proposition}[section]

\newtheorem{Theorem}{Theorem}[section]
\newtheorem{Lemma}{Lemma}[section]
\newtheorem{Corrolary}{Corrolary}[section]

\newcommand{\bp}{\begin{Proposition}}
  \newcommand{\ep}{\end{Proposition}}
\newcommand{\bt}{\begin{Theorem}} \newcommand{\et}{\end{Theorem}}
\newcommand{\bl}{\begin{Lemma}} \newcommand{\el}{\end{Lemma}}
\newcommand{\bc}{\begin{Corrolary}} \newcommand{\ec}{\end{Corrolary}}

  \def\om{\omega}
   \def\ep{\eps}
 \def\hn{{ \hat N}} \def\hq{{\hat Q}} \def\hp{{\hat
    \Phi}} \def\hs{{ \hat S}} \def\ha{{\hat A}} \def\cs{{\cal S}}

\author{K. Landsteiner\\Instituto de F{\'\i}sica Te\'orica
  C-XVI\\Universidad Aut{\'o}noma de Madrid\\28049
  Madrid,Spain\\Karl.Landsteiner@uam.es}

\author{C. I.  Lazaroiu\\Humboldt Universit\"at zu Berlin\\
  Newtonstrasse 15, 12489
  Berlin-Adlershof\\Germany\\calin@physik.hu-berlin.de}

\author{Radu Tatar\\Theoretical Physics Group \\Lawrence Berkeley
  National Laboratory\\Berkeley, CA 94720,
  USA\\rtatar@socrates.berkeley.edu }

\title{Chiral field theories, Konishi anomalies and matrix models}

\abstract{We study a chiral ${\cal N}=1$, $U(N)$ field theory in the
  context of the Dijkgraaf-Vafa correspondence. Our model contains one
  adjoint, one conjugate symmetric and one antisymmetric chiral multiplet, as
  well as eight fundamentals.  We compute the generalized
  Konishi anomalies and compare the chiral ring relations they induce 
  with the loop equations of the (intrinsically holomorphic) 
  matrix model defined by the tree-level superpotential of the field
  theory. Surprisingly, we find that the matrix model is well-defined 
  only if the number of flavors equals two!  Despite this mismatch, we
  show that the $1/{\hat N}$ expansion of the loop equations agrees with the generalized 
  Konishi constraints. This
  indicates that the matrix model --- gauge theory correspondence should
  generally be modified when applied to theories with net chirality.
  We also show that this chiral theory produces the same gaugino
  superpotential as a nonchiral $SO(N)$ model with a single symmetric
  multiplet and a polynomial superpotential. }

\preprint{HU-EP-03/29 \\
  IFT-UAM/CSIC-03-16 \\
  UCB-PTH-03/13;~LBNL-53386}

\begin{document}

\tableofcontents

\pagebreak

\vskip .6in

\section{Introduction}
\label{intro}

A surprising feature of ${\cal N}=1$ strong coupling dynamics was
uncovered in the seminal work of Dijkgraaf and Vafa \cite{DV,DV2,
  DV3}, who found a relation between the gaugino superpotential of a
confining ${\cal N}=1$ theory and certain holomorphic \cite{holo}
matrix models.  The recipe they proposed takes the tree-level
superpotential of such a theory to be the ´action´ of the dual matrix
model. This conjecture was proved for a few nontrivial examples, via
two distinct methods.  One approach \cite{dglvz} uses covariant
superfield techniques in perturbation theory to integrate out massive
matter fields in a gaugino background.  A different method was
proposed in \cite{Gorski, cdsw}, where it was shown that the loop
equations of the matrix model coincide formally with chiral ring
relations induced by certain generalizations of the Konishi anomaly.

Up to now the matrix model -- field theory correspondence has been
applied almost exclusively to the non-chiral case \footnote{See
  however \cite{Brandhuber, ferreti} for calculations of the effective
  superpotential using Konishi anomaly relations in certain chiral
  field theories.} \cite{DV}-\cite{cachazo}. The motivation of the
present paper is to test the conjecture for the case of chiral models.
Ideally, one would like to know if matrix models can be used to
calculate effective superpotentials of SUSY-GUTs or other
supersymmetric extensions of the standard model. Of course we are far
from answering this question.  Instead, we will study a model with
gauge group $U(N)$ and chiral matter content chosen to allow for a
straightforward large N limit. The matter consists of a field $\Phi$
in the adjoint representation, two fields $A$, $S$ in the
antisymmetric and conjugate symmetric two-tensor representations, and
eight fields $Q_1\dots Q_8$ in the fundamental representation to
cancel the chiral anomaly.  The tree level superpotential has the
form:
\begin{equation}
  \label{eq:superpot}
  W_{\mathrm{tree}} = \tr [ W(\Phi) + S \Phi A] + 
\sum_{f=1}^{8} Q_f^T S Q_f~~, 
\end{equation}
where $W$ is a complex polynomial.

This model has the advantage that the number of matter fields is
independent of the rank of the gauge group, thus allowing for a large
N limit with fixed matter content.  Further motivation to study this
model is provided by its interesting type IIA/M-theory realization
\cite{Karl_chiral}-\cite{park} \footnote{We note that there also
  exist IIB brane configurations describing chiral models, for example
  the so-called ``brane boxes'' of \cite{hz}, whose stability is
  unfortunately unclear. Geometric duals for these configurations are
  not currently known. }.  By taking the strong coupling limit (which
amounts to lifting the brane configuration to M-theory), it was argued
in these references that the model is described by a smooth curve.  This
can be interpreted as the existence of a mass gap and confinement, a
conclusion which is of course also suggested by our model's one loop
beta function. The geometric engineering of such models is discussed
in \cite{chiral2} by using methods of \cite{llt},
\cite{OT}-\cite{Cachazo_Vafa_more}.

In the present paper, we study the gaugino superpotential obtained
after confinement. We shall show that the effective
superpotential agrees with that of a different ${\cal N}=1$ theory,
namely a non-chiral $SO(N)$ model with a single chiral superfield $X$
in the symmetric representation and a tree-level superpotential given by
$\tr W(X)$ (the gaugino superpotential for such models was recently
investigated in \cite{KRS}).  This relation can be understood most
easily by turning on a D-term deformation of the original model, under
which the theory flows at low energies to the $SO(N)$ model with
symmetric matter. Because the effective superpotential is protected by
holomorphy, its form must be independent of the choice of
Fayet-Iliopoulos term, which explains why one obtains agreement
between the two theories. One can make this argument more
precise by computing the first (a.k.a. Veneziano-Yankielowicz)
approximation to the gaugino superpotentials upon using scale
matching techniques, and we shall do so below, finding agreement. To
give a complete proof of low-energy equivalence, we use the more
powerful technique of chiral ring relations \cite{cdsw}, which allows
us to characterize the exact effective superpotential in terms of 
solutions to certain algebro-differential equations induced by
generalized Konishi anomaly constraints. Then the connection between
the chiral and $SO(N)$ models follows upon matching the relevant
relations in the chiral rings.

In principle, the generalized Konishi constraints derived below
suffice to completely determine the exact gaugino superpotential,
which can be extracted with arbitrary precision by solving the relevant
equations. However, it is interesting to follow the beautiful insight
of Dijkgraaf and Vafa in order to construct a holomorphic matrix model
whose free energy specifies the superpotential. This can be achieved
by building a matrix integral whose loop equations reproduce the Konishi
constraints. Applying these ideas, one 
finds some novel phenomena, which are related to the chiral
character of our matter representation.

In fact, the holomorphic matrix model with action given by the tree level superpotential of our
chiral field theory turns out to be ill defined. The problem is that,
although the matrix model action is invariant under the
complexified $GL({\hat N},\C)$  gauge group, the measure fails 
to be invariant unless the number of matrix model flavors ${\hat Q}_f$ equals two.  This phenomenon,
which is due to the presence of a chiral matter content, 
forces us to work with a matrix model which contains only two flavors, even though the
associated field theory contains eight !  
Then the matrix partition function is well-defined, and we show that the loop
equations agree with the generalized Konishi constraints despite
the mismatch in the number of flavors. Our matrix model is intrinsically holomorphic, 
in the sense that it does not admit a real or Hermitian version. This is due to the fact that 
our matter representation is chiral. 

Having extracted the relevant matrix model, we compare
it with the model which governs the gaugino superpotential of
the $SO(N)$ theory with symmetric matter, a comparison which sheds
different light on the relation between the two theories.

The paper is organized as follows.  In Section \ref{fieldtheory}, we
analyze the classical moduli space of our theory by solving the F- and
D-flatness constraints. For a diagonal vev of $\Phi$, we find that the
gauge group is broken to a product $\prod_{i=1}^d U(N_i)$, where each
factor contains the same massless matter as the original theory.
Computation of the leading contribution to the effective
superpotential requires threshold matching, which cannot be performed
directly since chirality forbids the addition of a mass term.  This
problem was also encountered in \cite{Brandhuber}, where it
was solved by deforming the superpotential in such a way as to Higgs
the gauge group.  We will use a similar technique. Instead of
deforming the tree-level superpotential, use independence of the effective
superpotential of D-term deformations, which allows us to add a
Fayet-Iliopoulos term. In the presence of a Fayet-Iliopoulos parameter
(which we take to be positive), we find that the symmetric field
acquires a vev, which breaks the gauge group down to $SO(N)$ with
massless matter $X$ transforming in the symmetric representation and a
superpotential $\tr W(X)$. Then the first approximation to the gaugino
superpotential can be computed by standard threshold matching. The
resulting nonchiral $SO(N)$ model was recently studied in \cite{KRS} via the
Konishi anomaly approach of \cite{cdsw}.

In Section \ref{konishi}, we derive the generalized Konishi
constraints for our chiral theory and show how they relate to those extracted
in \cite{KRS} for the $SO(N)$ model. Since our
chiral model has a different matter content, we find a different set
of resolvent-like objects which enter the relevant `loop equations'.
However, we show that all such quantities are uniquely determined by
the solution of a pair of equations which coincide with those derived
in \cite{KRS} for the $SO(N)$ theory with symmetric matter.

Section \ref{matrixmodel} discusses the matrix model dual to our chiral theory. We show
that the measure is not invariant under the central $\C^*$ of the $GL({\hat N},\C)$
gauge group (and thus the matrix model partition function vanishes or is infinite) 
unless the number ${\hat N}_F$ of matrix model flavors equals two.  
We then extract the loop equations of this model by using both the
standard method of the eigenvalue representation and the approach of \cite{Naculich}. 
Finally, we discuss the relation with
the matrix integral relevant for the $SO(N)$ theory with symmetric
matter.

The identifications mapping our loop equations into the
Konishi constraints are given in section \ref{matrix.vs.field}. Using 
this map, we extract an explicit formula expressing the gaugino superpotential
in terms of the matrix model free energy. This completes the proof of
the Dijkgraaf-Vafa conjecture for our case.

Section \ref{conclusion} presents our conclusions. 
In appendix \ref{measure_inv} we prove gauge-invariance of the matrix model measure. 
Appendix \ref{SO_moduli} recalls the classical vacua of the $SO(N)$
theory with symmetric matter, while appendix \ref{konishi_proofs}
contains some details relevant for the discussion of Section
\ref{konishi}.

\section{A first view of field theory properties}
\label{fieldtheory}

In this section we take a first look at our field theory model. After
describing it precisely, we discuss the part of the classical moduli
space which will be relevant for our purpose, and give our
derivation of the Veneziano-Yankielowicz superpotential, which is the
first approximation to the exact glueball superpotential predicted by
the Dijkgraaf-Vafa correspondence.

\subsection{Description of the model}
\label{description}

We start with a $U(N)$ gauge group, together with chiral matter $\Phi,
S, A$ in the adjoint, antisymmetric and conjugate symmetric
representations, as well as $N_F$ quarks $Q_f$ in the fundamental
representation. We consider the tree-level superpotential:
\begin{equation}
\label{W_tree}
W_{tree}=\tr \left[ W(\Phi) + S  \Phi A\right] + \sum_{f=1}^{N_F}{Q_f^T SQ_f}~~,
\end{equation}
where:
\begin{equation}
\label{W}
W(z)=\sum_{j=1}^{d+1}{\frac{t_j}{j}z^j}
\end{equation}
is a complex polynomial of degree $d+1$. We have $S^T=S, A^T=-A$
(while $\Phi$ is unconstrained) and the gauge transformations are:
\begin{equation}
\label{field_gauge}
\Phi\rightarrow U {\hat \Phi} U^\dagger~~,~~S\rightarrow
{\bar U} S U^\dagger~~,~~A\rightarrow U A
U^T~~,~~Q_f\rightarrow UQ_f~~,
\end{equation}
where $U$ is valued in $U(N)$. The $U(N)$ gauge symmetry is obviously
preserved by $W_{tree}$. Note that the fields $S,A$ are {\em complex}.

We do not have quarks in the anti-fundamental representation and
therefore this system is quite different from models studied in
\cite{Seiberg, Radu_fundamentals1, Radu_fundamentals2, Hofman}. In
fact, the matter representation is chiral, in contrast to most
situations previously studied in the context of the Dijkgraaf-Vafa
conjecture.  In particular, the model will have a chiral anomaly
unless we take $N_F=8$. In the following, we shall focus on the
non-anomalous case though we allow $N_F$ to take an arbitrary value in
most formulas (this permits us to recover the anomaly cancellation
constraint $N_F=8$ as a consistency condition required by gaugino
condensation).

This model can be obtained through an orientifolded Hanany-Witten
construction \cite{Karl_chiral,hana,kutasov}. It can also be realized
through geometric engineering, as we discuss in a companion paper
\cite{chiral2}.

\subsection{The classical moduli space}
\label{modulispace}

Let us study the classical moduli space of our theories.  Part of the
discussion below is reminiscent of that given in
\cite{OT,Cachazo_Vafa} and \cite{us} for quiver gauge theories, though
of course we have a rather different matter content and hence the
details are not the same.

The F-flatness constraints are:
\begin{equation}
\label{eq1}
\Phi^{T} S= S \Phi~~,
\end{equation}
\begin{equation}
\label{eq2}
\Phi A - A \Phi^{T} + 2 \sum_{f=1}^{N_F}{Q_f Q_f^T}= 0~~,
\end{equation}
\begin{equation}
\label{eq3}
W'(\Phi) + A S = 0~~,
\end{equation}
\begin{equation}
\label{eq4}
S Q_f = 0~~,
\end{equation}
while the D-flatness condition is:
\begin{equation}
\label{Dflatness}
 \frac{1}{2} [\Phi^\dagger, \Phi]+S^\dagger S -AA^\dagger -
\frac{1}{2}\sum_{f=1}^{N_F}{Q_fQ_f^\dagger}=0~~, 
\end{equation}
where the left hand side is the moment map for our representation of
$U(N)$.

To understand the solutions, notice that (\ref{eq3}) implies:
\begin{equation} W'(\Phi)^2=(AS)^2=ASAS~~.  \end{equation}Using the
transpose $SA=W'(\Phi^T)$ of (\ref{eq3}) in the right hand side gives:
\begin{equation}
\label{Wintmd}
W'(\Phi)^2=AW'(\Phi^T)S=AS W'(\Phi)
\end{equation}
where in the last equality we used equation (\ref{eq1}). Applying
(\ref{eq3}) once again in the right hand side of (\ref{Wintmd}), we
find:
\begin{equation}
\label{crit}
W'(\Phi)^2=0~~.
\end{equation}

Let us assume that $[\Phi^\dagger, \Phi]=0$, i.e. $\Phi$ is a normal
matrix.  Then $\Phi$ is diagonalizable via a unitary gauge
transformation, and equation (\ref{crit}) shows that $\Phi$ can be
brought to the form:
\begin{equation}
\label{Phi_diag}
\Phi={\rm diag}(\lambda_1 {\bf 1}_{N_1}\dots \lambda_d {\bf 1}_{N_d})~~
\end{equation}
where $\lambda_1\dots \lambda_d$ are the distinct roots of $W'(z)$ and
$N_1\dots N_d$ and non-negative integers such that $N_1+\dots +N_d=N$.
If $N_j=0$ for some root $\lambda_j$, we use the convention that the
corresponding block $\lambda_j {\bf 1}_{N_j}$ does not appear in
(\ref{Phi_diag}).

With this form of $\Phi$, equation (\ref{eq1}) shows that $S$ must be
block-diagonal:
\begin{equation}
\label{S_bdiag}
S={\rm diag}(S_1\dots S_d)
\end{equation}
where $S_j$ are symmetric $N_j\times N_j$ matrices.  When bringing
$\Phi$ to the form (\ref{Phi_diag}), we are left with a residual
$\prod_{j=1}^d{U(N_j)}$ gauge symmetry corresponding to the
transformations $U={\rm diag}(U_1\dots U_d)$ with $U_j\in U(N_j)$.
Using this symmetry, we can bring $S_j$ to the form\footnote{This is
  the Takagi factorization (see, for example, \cite{Horn} page 204)
  for the complex symmetric matrix $S_j$.}:
\begin{equation}
\label{S_diag}
S_j={\rm diag}(0_{N_j^{(0)}}, \sigma_j^{(1)} {\bf 1}_{N_j^{(1)}}\dots
\sigma_j^{(m_j)} {\bf 1}_{N_j^{(m_j)}})~~
\end{equation}
where:
\begin{equation}
0=\sigma_j^{(0)}<\sigma_j^{(1)} < \dots <\sigma_j^{(m_j)}~~. 
\end{equation}
The multiplicities $N_j^{(k)}$ satisfy
$\sum_{k=0}^{m_j}{N_j^{(k)}}=N_j$.  In the case $m_j=0$, we have
$N_j^{(0)}=N_j$ and equation (\ref{S_diag}) reduces to $S_j=0_{N_j}$.
After bringing $S$ to the form (\ref{S_diag}), we are left with the
gauge symmetry $\prod_{j=1}^d\left[U(N_j^{(0)})\prod_{k=1}^{m_j}
  {SO(N_j^{(k)})}\right]$.

Writing:
\begin{equation}
\label{Q_vector}
Q_f=\left[\begin{array}{c}Q_f^{(1)}\\ \vdots
    \\Q_f^{(d)}\end{array}\right]
\end{equation}
where $Q_f^{(j)}$ is an $N_j$-vector, equation (\ref{eq4}) requires
that each $Q_f^{(j)}$ lies in the kernel of $S_j$. Thus we must have:
\begin{equation}
\label{Q_j}
Q_f^{(j)}=\left[\begin{array}{c}q_f^{(j)}\\0\\ \vdots
    \\0\end{array}\right]
\end{equation}
when decomposing into sub-vectors according to
$N_j=N_j^{(0)}+N_j^{(1)}+\dots +N_j^{(m_j)}$. Here $q_f^{(j)}$ is a
column $N_j^{(0)}$-vector.  Using the remaining gauge symmetry, we
bring the $N_j^{(0)}\times N_F$ matrix $q^{(j)}:=\left[q_1^{(j)}\dots
  q_{N_F}^{(j)}\right]$ to the form:
\begin{equation}
\label{q_j}
q^{(j)}:=\left[\begin{array}{ccccccc} 
    a_1&*&\dots &*&*&\dots &*\\0&a_2&\dots&*&*&\dots &*\\
    \vdots &\vdots&\ddots&\vdots&\vdots&\vdots&\vdots\\
    0&0&\dots &a_{s_j}&*&\dots &*\\0&0&\dots &0&0&\dots&0\\
    \vdots &\vdots&\vdots&\vdots&\vdots&\vdots&\vdots\\
    0&0&\dots &0&0&\dots&0 \end{array}\right]~~,
\end{equation}
where $0<a_1<\dots <a_{s_j}$, the symbol $*$ stands for generally
distinct complex entries and the zero rows at the bottom may be
absent. The rank $s_j$ of this matrix equals the dimension of the
vector space spanned by $q_1^{(j)}\dots q_{N_F}^{(j)}$, and can take
values between $0$ and ${\rm min}(N_j^{(0)},N_F)$. If we let
$p_j:=N_j^{(0)}-s_j$, then the vevs (\ref{q_j}) break the
$U(N_j^{(0)})$ components of the gauge group down to $U(p_j)$.

Finally, equation (\ref{eq2}) shows that in such a vacuum the
$N_i\times N_j$ blocks of $A$ are given by:
\begin{equation}
\label{A_vev}
A_{ij}=\frac{2}{\lambda_j-\lambda_i}\sum_{f=1}^{N_F}{Q_f^{(i)}(Q_f^{(j)})^T}~~.
\end{equation}
Since $Q^{(i)}_f$ lie in the kernel of $S_i$, this automatically
satisfies condition (\ref{eq3}), which in our vacuum takes the form
$AS=0\Longleftrightarrow SA=0$. Equation (\ref{A_vev}) shows that all
non-vanishing entries of $A$ lie in the $s_i\times s_j$ sub-blocks of
the $N_i^{(0)}\times N_j^{(0)}$ blocks, where they are given by
equations (\ref{A_vev}) and (\ref{q_j}).

Since we assume $[\Phi^\dagger,\Phi]=0$, the D-flatness condition
(\ref{Dflatness}) reduces to:
\begin{equation}
\label{Dflatness0}
S^\dagger S -AA^\dagger -
\frac{1}{2}\sum_{f=1}^{N_F}{Q_fQ_f^\dagger}=0~~.  \end{equation}Using the form
of $\Phi$, $S$ and $A$ discussed above (which is required by
F-flatness), it is not hard to see that (\ref{Dflatness}) implies
(note that we do {\em not} sum over $i$): \begin{equation}
\label{S_ivanish}
S_i^\dagger S_i=0~~{\rm for~all~} i ~~.
\end{equation}
This follows by decomposing (\ref{Dflatness0}) into $N_i\times N_j$
blocks and restricting to the case $i=j$ while using the fact that
$(AA^\dagger)_{ij}=-(A{\bar A})_{ij}=-A_{ik}{\bar A_{kj}} $ and
$(Q_fQ_f^\dagger)_{ij}=Q_f^i{\bar Q_f^j}$ both vanish for $i=j$ due to
equations (\ref{Q_vector}) and (\ref{A_vev}) except in the
$N_i^{(0)}\times N_i^{(0)}$ block. This implies that the entries of
the block diagonal matrix $S_i S^\dagger_i$ vanish except in this
block. The entries also vanish in the $N_i^{(0)}\times N_i^{(0)}$
block because of the form (\ref{S_diag}).

Since $S_i^\dagger S_i$ is positive semidefinite, equations
(\ref{S_ivanish}) imply $S_i=0$ for all $i$ and thus $S=0$.  In this
case, the D-flatness condition (\ref{Dflatness0}) becomes:
\begin{equation}
\label{Dflatness1}
AA^\dagger + \frac{1}{2}\sum_{f=1}^{N_F}{Q_fQ_f^\dagger}=0~~,
\end{equation}
which implies $A=0$ and $Q_f=0$ for all $f$ by semi-positivity of the
left hand side.

It follows that the only classical vacua for which $[\Phi^\dagger,
\Phi]=0$ are given by $\Phi$ of the form (\ref{Phi_diag}) and $S=A=0$
as well as $Q_f=0$ for all $f$. In such a vacuum, the gauge group is
broken down to the product $\prod_{j=1}^d{U(N_j)}$.

\subsection{The Veneziano-Yankielowicz superpotential}
\label{VY_section}

We next discuss the leading approximation to the gaugino
superpotential.  As we shall see below, the effective superpotential
coincides with that of an $SO(N)$ field theory with a single chiral
superfield $X$ transforming in the symmetric two-tensor
representation, and a tree-level superpotential $\tr W(X)$.

To compute the gaugino superpotential, we need the scale(s) of the low
energy theory, which are usually obtained via threshold matching.
Standard threshold matching is difficult to apply to chiral theories,
since a chiral tree-level action cannot contain mass terms for the
chiral fields, and thus one cannot directly integrate out such fields
at one-loop.  This problem was encountered for a chiral model
considered in \cite{Brandhuber}, where it was overcome by applying
threshold matching to a certain Higgs branch.  Some aspects of the
same issue were recently discussed in \cite{nakayama}.

It turns out that threshold matching can be carried out in our case
provided that one first deforms the theory through the addition of a
Fayet-Iliopoulos term $\xi$ (this is useful for our purpose since
holomorphy dictates that the low energy scale is insensitive to D-term
deformations).  Such a deformation was previously considered in
\cite{Karl_chiral,hana,kutasov} and has the following effect.
Concentrating on vacua with $[\langle \Phi\rangle, \langle
\Phi\rangle^\dagger]=0$, it is not hard to see that the inhomogeneous
form of (\ref{Dflatness}) (obtained by introducing $\xi {\bf 1}_N$ in the
right hand side) together with the F-flatness constraints again imply
$\langle A \rangle=0$ and $\langle Q_f\rangle =0$ for all $f$. On the
other hand, the symmetric field $S$ gets an expectation value equal to
the square root of $\xi$, which breaks $U(N)$ to $SO(N)$. The vev of
$S$ gives equal masses to the quarks $Q_{f}$ due to the terms
$Q_f^TSQ_f$.  The term $\tr S \Phi A$ gives equal masses to the fields
$A$ and $Y$ where $Y=\frac{1}{2}(\Phi-\Phi^T)$ is the antisymmetric
part of $\Phi$. All of these masses depend on the FI-term. The
symmetric part $X:=\frac{1}{2}(\Phi+\Phi^T)$ of the adjoint field
remains massless. The fluctuations of $S$ around its vev are 'eaten
up' while giving masses to the appropriate W-bosons through the Higgs
mechanism.

Eventually, one is left with an $SO(N)$ theory with a symmetric tensor
$X$ and a superpotential $\tr W(X)$.  By holomorphy, the scale of this
$SO(N)$ theory must be independent of the original FI-parameter. It is
easy to check this explicitly. Let the FI-parameter be $\xi=\nu^2$ for
some real $\nu$. Then the inhomogeneous form of (\ref{Dflatness})
shows that the vev of $S$ equals $\pm \nu$, and we can take the plus
sign without loss of generality. The massive fields are eight quark
flavors, two antisymmetric tensors ($A$ and the antisymmetric part of
$\Phi$) and the W-bosons in the symmetric representation of the low
energy gauge group $SO(N)$. This gives the scale matching relation:
\begin{equation}\label{scalematching.one}
\Lambda_0^{N-4} = \Lambda^{N-4} \nu^4 \nu^{N-2}
\nu^{-(N+2)}=\Lambda^{N-4}\Longrightarrow \Lambda_0=\Lambda ~~.
\end{equation}
Here $\Lambda$ is the scale of the chiral high energy theory (whose
one-loop beta function coefficient is $N-4$) and $\Lambda_0$ is the
scale of the low energy $SO(N)$ theory. Notice that the exponent of
$\Lambda_0$ is unusual in the sense that we would expect $2N-8$ for
the $SO(N)$ theory with symmetric tensor. This can be traced back to
the fact that the generators in the $SO(N)$ theory are unusually
normalized. It can be seen by noticing that the index for the
fundamental representation of $U(N)$ is $\frac 1 2$ whereas the index
for the fundamental representation of $SO(N)$ with conventional
normalization is $1$ and that a fundamental of the high energy $U(N)$
theory descends directly to a fundamental of the low energy $SO(N)$
theory. This normalization has already been taken into account in
(\ref{scalematching.one}).

We still have the deformation $\tr W(X)$, which leads to diagonal vevs of
$X$ of the form (\ref{Phi_diag}). The relevant vacua of the $SO(N)$
theory with a symmetric field are discussed in Appendix
\ref{SO_moduli}.  One finds that the vev of $X$ further breaks the Lie algebra of the low
energy gauge group according to \cite{intriligator}:
\begin{equation}
so(N)\rightarrow \oplus_{i=1}^{d} so(N_i)~~. 
\end{equation}
It is now easy to extract the scales of the different $so(N_i)$
factors (here we use conventional normalization, since we compare only $SO$
theories):
\begin{equation}
\label{scales}
\Lambda_{i}^{3(N_i - 2)} =
\left[m_{X_i}^{N_i + 2} \prod_{j\neq i}\, m_{W_{ij}}^{- 2 N_j}\right] \Lambda^{2(N
  - 4)} 
\end{equation}
where $ m_{X_i} = W''(\lambda_i)$ and the masses of the
$SO(N)/SO(N_i)$ W-bosons are $m_{W_{ij}}=\lambda_i - \lambda_j$.
Because of (\ref{scalematching.one}), we can use the high energy scale
$\Lambda$ in (\ref{scales}).

The Veneziano-Yankielowicz contribution to effective superpotential
has the form:
\begin{equation}
\label{VYpot}
W_{eff} =
\sum_{i=1}^{d} \frac{\cs_i}{2} [\log \left(\frac{\Lambda_{i}^{3(N_i -
      2)}}{\cs_i^{N_i - 2}}\right) + N_i - 2] \,,
\end{equation}
where we have inserted the factor $1/2$ coming from the relative
normalization of the generators of $U(N)$ and $SO(N)$.

\paragraph{Observation:}
 
One can also consider turning on a negative FI-term $\xi=-\nu^2$.
Taking $N$ to be even for simplicity and assuming $ [\langle
\Phi\rangle^\dagger, \langle \Phi\rangle ]=0$ and $\langle
Q_f\rangle=0$ for all $f$, the D-flatness condition shows that the
antisymmetric field $A$ acquires a vev. Up to a gauge transformation,
we can take $\langle A \rangle = \nu J$ where $J$ is the antisymmetric
invariant tensor of $Sp(N/2)$:
\begin{equation}
J = \left( 
  \begin{array}{cc}
0& {\bf 1}_{N/2}\\
-{\bf 1}_{N/2}&0
  \end{array}
\right)  \,.
\end{equation}
This breaks the $U(N)$ gauge group down to $Sp(N/2)$, with the
fluctuations of $A$ 'eaten' by the W-bosons of the coset
$U(N)/Sp(N/2)$. Let us decompose $\Phi=\Phi_+ + \Phi_-$,
where $(\Phi_\pm J)^T=\pm (\Phi_\pm J)$. Then the superpotential
becomes:
\begin{equation}
  \label{eq:spbreaking}
 \tr[ W(\Phi_+ + \Phi_-) +\nu S\Phi_+J + Q_f\otimes Q_f^T S ]~~.  
\end{equation}
Integrating out $\Phi_+$ by the equation of motion of $S$ (which gives
$\Phi_+= -\frac{1}{\nu} Q_f\otimes Q_f^T J$) results in the
superpotential:
\begin{equation}
  \label{eq:spsuperpot}
W_{low}=W(\Phi_- -\frac{1}{\nu} Q_f\otimes Q_f^T J)~~.
\end{equation}
Hence at low energies we have an $Sp(N/2)$ gauge theory with an
antisymmetric tensor and eight fundamentals interacting through the
superpotential (\ref{eq:spsuperpot}).  Because of the complicated
structure of $W_{low}$, we did not find this branch to be useful for
our purpose.

\section{Low energy analysis via generalized Konishi anomalies}
\label{konishi}

In this section, we extract the relevant chiral ring relations of our
model and compare with those of the $SO(N)$ theory with
symmetric matter. We shall use the method of generalized Konishi
anomalies originally developed in \cite{Gorski, cdsw}. The structure of
the tree level superpotential implies:
\begin{equation}
\label{Wj}
  j \frac{\partial W_{eff}}{\partial t_j} = \langle \tr(\Phi^j)\rangle~~.
\end{equation}
Our strategy is to extract a set
of Konishi anomaly relations  which allow one to solve for the generating
function $T(z) = \langle \tr(\frac{1}{z-\Phi})\rangle$ of the chiral
correlators $\langle \tr(\Phi^j)\rangle$ appearing in the right hand side.
The integration of (\ref{Wj}) allows one to compute the
effective superpotential up to a piece which is independent of the
coupling constants $t_j$.

\subsection{Konishi constraints for the chiral model}
\label{konishiconstraints}

As discussed in \cite{cdsw}, the loop equations of the (adjoint) one-matrix
model are formally equivalent to certain chiral ring relations induced
by a generalized form of the Konishi anomaly of the $U(N)$ field
theory with one adjoint multiplet.  The argument extends to other
matter representations, and is based on special properties of chiral
operators (=operators corresponding to the lowest component of chiral
superfields) in ${\cal N}=1$ supersymmetric field theories in four
dimensions.  It is well-known that correlators of such operators do
not depend on the space-time coordinates and that they factorize:
\begin{equation}
\langle  {\cal O}_1  {\cal O}_2 \rangle = \langle {\cal O}_1 \rangle \langle  
{\cal O}_2 \rangle~~.
\end{equation}
On the set of chiral operators one considers the equivalence relation:
\begin{equation}
\label{equiv}
 {\cal O}_1 \equiv {\cal O}_2 + c_{\dot{\alpha}} \bar{Q}^{\dot{\alpha}} (\dots)~~,
\end{equation}
where $c_{\dot{\alpha}} \bar{Q}^{\dot{\alpha}} $ is an arbitrary
linear combination of the anti-chiral supercharges.
This equivalence relation is compatible with the operator
product structure, and modding out the space of chiral operators by
(\ref{equiv}) leads to the chiral ring.  Equivalence of two chiral
operators under (\ref{equiv}) implies equality of their vevs:
\begin{equation}
\left\langle  {\cal O}_1\right\rangle = \left\langle  {\cal O}_2\right\rangle~~.
\end{equation}
An important relation in the chiral ring is:
\begin{equation}
\label{cr_w_alpha}
\{ {\cal W}_\alpha^{(r)} , {\cal W}_\beta^{(r)}\} \equiv 0~~,
\end{equation}
which holds for any representation $r$ of the gauge group.  This is a
special case of the general relation \cite{Brandhuber}:
\begin{equation}
{\cal W}_\alpha^{(r)}. {\cal O}^{(r)}\equiv 0~~,
\end{equation}
which holds for an arbitrary chiral operator ${\cal O}^{(r)}$
transforming in the representation $r$ of the gauge group.  The dot in
this equation indicates the action of ${\cal W}_\alpha$ in the
representation $r$.  For our theory, one finds:
\begin{eqnarray}\label{cr_phi}
{\cal W}_\alpha . \Phi &=& [{\cal W}_\alpha,\Phi] \equiv 0~~\\
{\cal W}_\alpha . A &=& {\cal W}_\alpha A + A {\cal W}_\alpha^T \equiv 0~~\\
{\cal W}_\alpha . S &=& - S {\cal W}_\alpha - {\cal W}_\alpha^T S \equiv 0~~\\
{\cal W}_\alpha . Q_f &=&{\cal W}_\alpha Q_f \equiv 0~~,\label{cr_q}
\end{eqnarray}
where ${\cal W}_\alpha$ in the right hand sides are taken in the adjoint representation and
juxtaposition stands for matrix multiplication. Note that ${\cal
  W}_\alpha$ acts on the product $A S$ through the commutator (since
this product transforms in the adjoint representation).

The generalized Konishi anomaly is the anomalous Ward identity for a
local holomorphic field transformation:
\begin{equation}
{\cal O}^{(r)} \longrightarrow {\cal O}^{(r)} + \delta {\cal O}^{(r)}~~.
\end{equation}
The supercurrent generator has the form $J = ({\bf O}^{(r)})^\dagger
e^{{\bf V}^{(r)}} \delta {\bf O}^{(r)}$ where ${\bf V}^{(r)}$ is the
vector superfield in the representation $r$ and ${\bf O}^{(r)}$ is the
chiral superfield associated with ${\cal O}^{(r)}$.  The chiral ring
relation induced by the generalized Konishi anomaly for this current
is:
\begin{equation}
\label{konishi_relation}
 \delta {\cal O}_I
\frac{\partial W}{\partial {\cal O}_I}  \equiv
- \frac{1}{32\pi^2}  {\cal W}^{\alpha}_I\,^J {\cal W}_{\alpha,J}\,^K
\frac{\partial (\delta {\cal O}_K)}{\partial {\cal O}_I }~~,
\end{equation}
where the capital indices enumerate a basis of the representation $r$.
We will investigate the generalized Konishi relations corresponding to
the field transformations:
\begin{eqnarray}
\delta \Phi &=& \frac{{\cal W}^\alpha {\cal W}_\alpha}{z-\Phi}  
\label{var.phi.one}\\
\delta \Phi &=& \frac{1}{z-\Phi} \label{var.phi.two}\\
\delta A &=& \frac{{\cal W}^\alpha}{z-\Phi}A 
\frac{({\cal W}_\alpha)^T}{z-\Phi^T}  \label{var.A.one}\\
\delta A &=& \frac{1}{z-\Phi}A \frac{1}{z-\Phi^T}  \label{var.A.two}\\
\delta S &=& \frac{1}{z-\Phi^T}S \frac{1}{z-\Phi}  \label{var.S.one}\\
\delta Q_f &=& \sum_{g=1}^{N_F}{\frac{\lambda_{fg}}{z-\Phi} Q_g}~~.
\label{var.Q}
\end{eqnarray}
In the last equation, $\lambda$ is an arbitrary matrix in flavor
space.

Writing ${\cal W}^2 = {\cal W}^\alpha {\cal W}_\alpha$, we define:
\begin{eqnarray}
{\bf R}(z) &:=& - \frac{1}{32\pi^2} \tr \left(\frac{{\cal W}^2}{z-\Phi}\right)~~\\
{\bf w}_\alpha(z) &:=& \frac{1}{4\pi} \tr \left(\frac{{\cal W}_\alpha}{z-\Phi}\right)~~\\
{\bf T}(z) &:=& \tr\left(\frac{1}{z-\Phi}\right)~~.
\end{eqnarray}
Then it is shown in Appendix \ref{konishi_proofs} that transformations
(\ref{var.phi.one}-\ref{var.Q}) generate the chiral ring relations:
\begin{eqnarray}
\label{konishi.phi.one}
-\frac{1}{32\pi^2}\tr \left( \frac{W'(\Phi){\cal W}^2}{z-\Phi} \right)
 -\frac{1}{32\pi^2}\tr\left( S \frac{{\cal W}^2}{z-\Phi} A \right) &\equiv&
{\bf R}(z)^2 ~~\\
\tr \left( \frac{W'(\Phi)}{z-\Phi}\right)
+  \tr \left(S \frac{1}{z-\Phi} A \right) &\equiv&
2 {\bf R}(z) {\bf T}(z) + {\bf w}^\alpha(z) {\bf w}_\alpha(z) ~~~~~~~~~~~~~~~\label{konishi.phi.two}~~\\
-\frac{1}{32\pi^2}\tr\left( S \frac{{\cal W}^2}{z-\Phi}A\right)&\equiv&
\frac{1}{2} {\bf R}(z)^2 \label{konishi.A.one}~~~~~~~~~~~~~~~~~~~\\
\tr\left(S\frac{1}{z-\Phi}A\right) &\equiv &
{\bf R}(z) {\bf T}(z) + 2 {\bf R}'(z) -
\frac{1}{2} {\bf w}^\alpha(z) {\bf w}_\alpha(z)\label{konishi.A.two}~~~~~~~~~~~~~~~`\\
 \tr\left( S \frac{1}{z-\Phi} A\right) +
\sum_{f} Q_f^T \frac{1}{z-\Phi^T} S \frac{1}{z-\Phi} Q_f  &\equiv&
{\bf R}(z) {\bf T}(z) - 2 {\bf R}'(z) - \frac{1}{2} {\bf w}^\alpha(z)
{\bf w}_\alpha(z)~~~~~~~~~~~~~~~\label{konishi.S.one}\\
2 Q_f^T S \frac{1}{z-\Phi} Q_g &\equiv& 
{\bf R}(z) \delta_{fg} ~~.~~~~~~~~~~\label{Qfg}
\end{eqnarray}
Taking the trace of the last equation gives:
\begin{equation}
2 Q_f^T S \frac{1}{z-\Phi} Q_f \equiv 
{\bf R}(z) N_F ~~.~~~~~~~~~~\label{konishi.Q.one}
\end{equation}

We next take the vacuum expectation values of these chiral ring
relations. Let us define:
\begin{eqnarray}
R(z) &:=&\langle {\bf R}(z)\rangle\label{R.konishi}\\
T(z)&:=&\langle {\bf T}(z)\rangle \label{T.konishi}\\
M(z) &:=& \left\langle \tr \left( S \frac{1}{z-\Phi}A\right)
\right\rangle~~\label{M.konishi}\\ 
M_Q(z) &=& \sum_f \left\langle
Q_f^T\frac{1}{z-\Phi^T}S\frac{1}{z-\Phi}Q_f \right\rangle~~
\label{MQ.konishi}\\ 
K(z) &:=& -\frac{1}{32\pi^2}\left\langle \tr
\left( S \frac{{\cal W}^2}{z-\Phi}A\right)
\right\rangle~~\label{K.konishi}\\ 
L(z) &:=& \sum_f \left\langle Q_f^T
S\frac{1}{z-\Phi}Q_f \right\rangle~~.
\label{L.konishi}
\end{eqnarray}
Introducing the degree $d-1$ polynomials:
\begin{eqnarray}
f(z) &=& -\frac{1}{32\pi^2} \tr\left( \frac{W'(z)-W'(\Phi)}{z-\Phi} {\cal W}^2 \right)
\label{fpoly.konishi}~~\\
c(z) &=&  \tr\left( \frac{W'(z)-W'(\Phi)}{z-\Phi}  \right) \label{cpoly.konishi}~~,
\end{eqnarray}
we write:
\begin{eqnarray}
-\frac{1}{32\pi^2}\tr \left( \frac{W'(\Phi){\cal W}^2 }{z-\Phi}\right) &=& 
W'(z) R(z) - f(z) ~~\\
\tr \left( \frac{W'(\Phi)}{z-\Phi}\right) &=& W'(z) T(z) - c(z) ~~.
\end{eqnarray}

Noticing that the vevs of spinor fields vanish due to Lorentz
invariance, we find the following Ward identities for the generating
functions (\ref{R.konishi}-\ref{L.konishi}):
\begin{eqnarray}
R(z)^2 - K(z) - W'(z) R(z) + f(z) & = & 0 \label{wi.konishi.R}\\
2 R(z) T(z) - W'(z) T(z) - M(z) + c(z) & = & 0 \label{wi.konishi.T}\\
K(z) - \frac{1}{2} R(z)^2 &=& 0 \label{wi.konishi.K} \\
M(z) - R(z) T(z) - 2 R'(z)  &=& 0 \label{wi.konishi.M}\\
M_Q(z) + M(z)  - R(z) T(z) + 2 R'(z)  &=& 0 \label{wi.konishi.MQ}\\
2L(z)- N_F R(z) &=& 0 \label{wi.konishi.L}~~. 
\end{eqnarray}

It is easy to eliminate $K(z), M(z), M_Q(z)$ and $L(z)$ from these
equations to find:
\begin{eqnarray}
R(z)^2 - 2 W'(z) R(z) + 2 f(z) & = & 0 \label{konishi.R}\\
T(z) R(z) - W'(z) T(z) - 2 R'(z) +  c(z) & = &0 \label{konishi.T}~~.
\end{eqnarray}
Given a solution $(R(z), T(z))$ of these constraints, the quantities
$K, M, M_Q$ and $L$ are given by:
\begin{eqnarray}
\label{konishi_sol}
K(z)&=&\frac{1}{2}R(z)^2\nn\\
M(z)&=&2R'(z)+R(z)T(z)\\
M_Q(z)&=&-4R'(z)\nn\\
L(z)&=&\frac{N_F}{2}R(z)~~.\nn
\end{eqnarray}
Hence all solutions are parameterized by the $2d$ complex coefficients
of the polynomials $f(z)$ and $c(z)$.

The generalized Konishi relations involving the flavors $Q_f$ have an
interesting implication. Expanding the last two equations in
(\ref{konishi_sol}) to leading order in $1/z$ gives:
\begin{equation}
\sum_f \langle Q_f^T S Q_f \rangle =4{\cal S}
\end{equation}
and
\begin{equation}
\sum_f \langle Q_f^T S Q_f \rangle=\frac{N_F}{2}{\cal S}~~,
\end{equation}
where ${\cal S}=-\frac{1}{32\pi^2} \langle \tr {\cal W}^2\rangle$ is
the gaugino condensate. If ${\cal S}$ is non-vanishing, then
compatibility of these two equations requires that we set $N_F=8$,
which is also required for canceling the chiral anomaly.  Any other
value is incompatible with the existence of a gaugino condensate.

\paragraph{Observation:}
The particular form of (\ref{Qfg}) is related to the $O(N_F)$ flavor
symmetry of the field theory with tree-level superpotential
(\ref{W_tree}):
\begin{equation}
Q_f\rightarrow Q_f':=r_{fg}Q_g~~,
\end{equation}
where $r$ is a general complex orthogonal matrix. Since this symmetry
is unbroken after confinement, the quantity $L_{fg}=\langle Q_f^T S
\frac{1}{z-\Phi} Q_g\rangle $ must be $O(N_F)$ invariant and thus
proportional to $\delta_{fg}$. Equation (\ref{Qfg}) shows that the
proportionality factor is given by $\frac{1}{2}R(z)$~~.

\subsection{Comparison with the $SO(N)$ model with symmetric matter}
\label{comparision}

The Konishi relations (\ref{konishi.R}) and (\ref{konishi.T}) coincide
with those of the $SO(N)$ field theory with a single complex chiral
superfield $X$ transforming in the symmetric representation, and with
a tree-level superpotential given by $\tr W(X)$.  The Konishi
relations for this theory were derived in \cite{KRS} (see equation
(18) of that reference\footnote{The paper \cite{KRS} considers a
  slight extension of our $SO(N)$ theory, by adding some fundamental
  and anti-fundamental matter beyond the symmetric field. The case of
  interest for our purpose is recovered from the equations of
  \cite{KRS} by setting their quark mass matrix to zero.}) and one
immediately checks that they agree with our equations
(\ref{konishi.R}) and (\ref{konishi.T}).  More precisely, the $SO(N)$
theory with symmetric matter admits the following resolvent-like
objects:
\begin{eqnarray}
R_{s}(z) &:=& -\langle \frac{1}{32\pi^2} \tr \left(\frac{{\cal W}^2}{z-X}\right)\rangle~~\\
T_{s}(z) &:=& \langle \tr\left(\frac{1}{z-X}\right)\rangle ~~,
\end{eqnarray}
which were shown in \cite{KRS} to obey the two equations
(\ref{konishi.R}) and (\ref{konishi.T}). This leads to the
identification:
\begin{eqnarray}
\label{field_ids}
R(z) &\longleftrightarrow& R_{s}(z)\nn\\
T(z) &\longleftrightarrow& T_{s}(z)~~.
\end{eqnarray}
which maps a solution of our Konishi constraints to a solution of the
Konishi relations of the $SO(N)$ theory (this works because equations
(\ref{konishi_sol}) completely determine the other resolvent-like
objects of the chiral theory given a solution $(R, T)$ of
(\ref{konishi.R}) and (\ref{konishi.T}) ).

\section{The matrix model}
\label{matrixmodel}

The general conjecture of \cite{DV} suggests that the effective superpotential of our field theory
should be described by the holomorphic matrix model\footnote{It is interesting to notice that this 
holomorphic matrix model does not have a real (`Hermitian') counterpart. This is due to the fact that our matter 
representation is intrinsically complex.}:
\begin{equation}
\label{Z_mod}
Z=\frac{1}{|G|}\int_\Gamma{d{\hat \Phi} d{\hat A} d{\hat S} d{\hat Q} e^{-\frac{\hn}{g}
{\cal S}_{mm}({\hat \Phi}, {\hat A}, {\hat S}, {\hat Q})}}~~,
\end{equation}
where $|G|$ is a normalization factor and:
\begin{equation}
\label{action}
{\cal S}_{mm}({\hat \Phi}, {\hat A}, {\hat S}, {\hat Q})=
\tr\left[ W({\hat \Phi})+{\hat S}{\hat \Phi} {\hat A} \right]+\sum_{f=1}^{{\hat N}_F}{{\hat Q}_f^T {\hat S} {\hat Q}_f}=
\tr\left[ W({\hat \Phi})+{\hat \Phi} {\hat A} {\hat S}+\sum_{f=1}^{{\hat N}_F}{{\hat Q}_f {\hat Q}_f^T {\hat S}} \right]~~.
\end{equation}
Here ${\hat \Phi}$ is an arbitrary complex ${\hat N}\times {\hat N}$
matrix, the complex ${\hat N}\times {\hat N}$ matrices ${\hat S}$ and
${\hat A}$ are symmetric and antisymmetric respectively and ${\hat Q}$
is a general complex ${\hat N}\times {\hat N}_F$ matrix whose columns we
denote by ${\hat Q}_f$.  The integration measure $d\mu= d{\hat \Phi}
d{\hat A} d{\hat S} d{\hat Q}=d{\hat \Phi} d{\hat A} d{\hat
  S}\prod_{f=1}^{{\hat N}_F} d{\hat Q}_f$ is given by:
\begin{equation}
\label{measure}
d{\hat \Phi}=\bigwedge_{i,j=1}^\hn{d{\hat \Phi}_{ij}}~~,
~~d{\hat S}=\bigwedge_{i\leq j}{d {\hat S}_{ij}}~~,~~d{\hat A}=\bigwedge_{i<j}{d{\hat A}_{ij}}~~,~~
d{\hat Q}=\bigwedge_{f=1}^{{\hat N}_F}{\bigwedge_{i=1}^{\hn}{d{\hat Q}_f^i}}~~.
\end{equation}
where $\bigwedge$ denotes the wedge product and we use the lexicographic order of 
indices to give unambiguous meaning to the various products of one-forms. 
For example, the notation 
$\bigwedge_{i\leq j}{d{\hat S}_{ij}}$ means:
\be
d\hs_{11}\wedge d\hs_{12}\wedge \dots \wedge d\hs_{1\hn}\wedge d\hs_{22}\wedge d\hs_{23}\wedge \dots \wedge d\hs_{2\hn}
\wedge \dots 
\wedge d\hs_{\hn-1\hn-1}\wedge d\hs_{\hn-1\hn}\wedge d\hs_{\hn\hn}~~.
\ee
Of course, the ordering convention can be chosen arbitrarily since changing it 
produces an irrelevant sign prefactor in the matrix integral.

The measure $d\mu$ is a top holomorphic form on the complex space:
\begin{equation}
{\cal M}=\{({\hat \Phi},{\hat S},{\hat A},{\hat Q}_1\dots {\hat
  Q}_{{\hat N}_F})|{\hat S}^T={\hat S}, {\hat A}^T=-{\hat A} \}~~.
\end{equation}
The integral in (\ref{Z_mod}) is performed on a boundary-less real
submanifold $\Gamma$ of ${\cal M}$ whose closure is non-compact and
which is chosen such that $\dim_\R \Gamma=\dim_\C {\cal M}$.  The model
(\ref{Z_mod}) admits the $O({\hat N}_F)$ flavor symmetry: \be
\label{flavor_m}
{\hat Q}_f\rightarrow {\hat Q}_f':=r_{fg}{\hat Q}_g~~,
\ee
where $r$ is an $\hn_F\times \hn_F$ orthogonal matrix~~.

Since anomaly cancellation in our field theory requires $N_F=8$, one is tempted 
to set ${\hat N}_F=8$ as well. It turns out that this naive identification cannot hold in our case. To
understand why, notice that both the matrix model action (\ref{action}) and the integration measure
are invariant under the following $SL(\hn,\C)$ gauge transformations:
\begin{equation}
\label{matrix_gauge}
{\hat \Phi}\rightarrow U {\hat \Phi} U^{-1}~~,~~{\hat S}\rightarrow
(U^{-1})^T {\hat S} U^{-1}~~,~~{\hat A}\rightarrow U {\hat A}
U^T~~,~~{\hat Q}_f\rightarrow U {\hat Q}_f~~,
\end{equation}
where $U$ is a complex $\hn\times \hn$ matrix of unit determinant (invariance of the measure is discussed in detail 
in Appendix \ref{measure_inv}).  To
preserve this symmetry, one must choose $\Gamma$ to be stabilized by
the action (\ref{matrix_gauge}) of $SL({\hat N},\C)$.

The matrix model action is in fact invariant under the full $GL({\hat
  N},\C)$ group acting as in (\ref{matrix_gauge}). However, the
measure $d\mu$ is {\em not} invariant under the central $\C^*$
subgroup of $GL({\hat N},\C)$ unless ${\hat N}_F=2$. Taking $U=\xi
{\bf 1}_{\hat N}$ in (\ref{matrix_gauge}) with $\xi\in \C^*$, we have:
\begin{equation}
\label{U1}
{\hat A}\rightarrow \xi^2 {\hat A}~~,~~{\hat S}\rightarrow
\xi^{-2}{\hat S}~~{\rm and}~~
{\hat Q}_f\rightarrow \xi {\hat Q}_f~~,
\end{equation}
which gives:
\begin{equation}
d\mu\rightarrow \xi^{{\hat N}({\hat N}_F-2)} d\mu~~.
\end{equation}

Let us assume ${\hat N}_F\neq 2$ and choose $\Gamma$ to be $GL(\hn,\C)$ invariant. 
Then, since the matrix model action
(\ref{action}) is invariant under (\ref{U1}), while the measure
transforms nontrivially, invariance of the integral (\ref{Z_mod}) under
coordinate transformations shows that:
\be
Z=\xi^{{\hat N}({\hat N}_F-2)} Z~~.
\ee 
Thus $Z$ must either vanish or equal complex infinity\footnote{The second solution is allowed since 
$Z$ is complex and the point at infinity in the complex plane does satisfy
$\infty= \xi^{{\hat N}({\hat N}_F-2)}\infty$. }! 
A similar argument shows that the integral $\int{d\mu F e^{-\frac{\hat N}g {\cal
      S}_{mm}}}$ must vanish  or be infinite for any functional $F(\hp,\ha,\hs,\{\hq_f\})$
which is invariant under (\ref{U1}). In particular, the expectation
value:
\begin{equation}
\langle F \rangle:=\frac{1}Z\int{d\mu F
  e^{-\frac{\hat N}{g} {\cal S}_{mm}}}
\end{equation}
of any such functional is ill-defined ! This
means that the matrix model predicted by a naive application of the
conjecture of \cite{DV} is not well-defined.

That subtleties can arise when
attempting to apply the conjecture of \cite{DV} to chiral field
theories is not completely unexpected, since most derivations of this conjecture
up to date have concentrated on real matter representations, which
prevent the appearance of net chirality. The phenomenon we just
discussed shows that one must modify the original conjecture of
\cite{DV} in order to adapt it to the chiral context.

Thus we are lead to consider the matrix model with ${\hat N}_F=2$.
Then both the action (\ref{action}) and the integration measure
are invariant under $GL(\hn,\C)$ transformations of the form
(\ref{matrix_gauge}), where $U$ is now an arbitrary complex invertible
matrix.  In Subsections \ref{eigenvaluerep} and \ref{relation} below, we shall show explicitly 
that the model with ${\hat N}_F=2$ is well-defined by relating it to the holomorphic matrix model
associated with the $SO(N)$ theory with symmetric matter.

\subsection{Loop equations}
\label{loopequations}

In this subsection, we extract the loop equations of the model (\ref{Z_mod}).
Although the correlation functions are not well defined unless
$\hn_F=2$, we will work formally with an arbitrary value of ${\hat N}_F$.  
This will allow us to re-discover the constraint ${\hat N}_F=2$ as a
consistency condition between the loop equations, in a manner similar to the way 
in which we recovered the condition $N_F=8$ in Subsection 3.1. by using the Konishi constraints of the field theory.

In addition to the matrix model resolvent:
\begin{equation}
\label{omega}
\omega(z) = \frac{g}{\hn} \tr \frac{1}{z-\hp} ~~,
\end{equation}
we shall consider the objects:
\begin{eqnarray}
k(z) &=& \frac{g}{\hn}\tr \left[ \hs \frac{1}{z-\hp} \ha \right]
\label{def.m}\\
m_Q(z) &=& \hq_f^T \frac{1}{z-\hp^T} \hs \frac{1}{z-\hp} \hq_f
\label{def.mQ}\\
l(z) &=& \hq_f^T \hs\frac{1}{z-\hp} \hq_f\label{def.l}~~.
\end{eqnarray}
We will show that these fulfill the loop equations:
\begin{eqnarray}
\label{loop_eqs}
\langle \om(z)^2 - W'(z) \om(z) - k(z) +\tilde f(z)\rangle &=&0~~\label{loop.om}\\
\langle \frac 1 2 \om(z)^2  + \frac 1 2 \frac{g}{\hat N} \om'(z)  - k(z)\rangle &=&0~~\label{loop.m} \\
\langle \om'(z)  +  m_Q(z) \rangle &=&0~~\label{loop.mq} ~~\\
\langle {\hat N}_F \omega(z)-2l(z) \rangle&=&0~~,~~\label{loop.l}
\end{eqnarray}
where:
\begin{equation}
\label{tilde_f}
\tilde{f}(z) := \frac{g}{\hn}\tr \frac{W'(z)-W'(\hp)}{z-\hp} ~~
\end{equation}
is a polynomial of degree $d-1$.

Before giving the derivation of these constraints, let us note that
one can eliminate $\langle k(z)\rangle$ between (\ref{loop.m}) and
(\ref{loop.om}) to find an equation for the resolvent:
\begin{equation}
\label{omega_loop}
\left\langle \om (z)^2 - \frac{g}{\hn} \om'(z) - 2 W'(z) \om(z)
+ 2 \tilde{f}(z) \right\rangle =0~~.
\end{equation}
Given a solution $\langle \omega(z)\rangle$, relation (\ref{loop.m}),
(\ref{loop.mq}) and (\ref{loop.l}) determine the averages of $k(z)$,
$m_Q(z)$ and $l(z)$ as follows:
\begin{eqnarray}
\label{loop_sol}
\langle k(z)\rangle &=&\frac{1}{2}\langle \omega(z)^2+\frac{g}{\hat N}\omega'(z)\rangle\nn\\
\langle m_Q(z)\rangle &=&-\langle \omega'(z)\rangle \\
\langle l(z)\rangle &=&\frac{{\hat N}_F}{2} \langle \omega(z)\rangle ~~.\nn
\end{eqnarray}
The leading order in the large $z$ expansion of the last two equations gives:
\begin{eqnarray}
  \label{eq:matrix.consisitency}
  \langle \hq_f^T \hs \hq_f \rangle &=& g~~\\
  \langle \hq_f^T \hs \hq_f \rangle &=& \frac{\hn_F}{2} g\,,
\end{eqnarray}
where we used the large $z$ behavior of the resolvent:
\be
\langle \om(z) \rangle
\approx \frac{g}{z} + O(\frac{1}{z^2})~~.
\ee
Since we of course take $g\neq 0$, equations (\ref{eq:matrix.consisitency}) are consistent only if ${\hat N}_F=2$.
We now proceed to give the proof of (\ref{loop.om}-\ref{loop.l}).

\subsection{Direct derivation of the loop equations}
\label{direct}

Consider the identity:
\begin{equation}
\int d\hp\, d\ha\, d\hs\, d{\hat Q}~
\frac{\partial}{\partial \hp_i\,^j} \left[
\left(\frac{1}{z-\hp}\right)_i\,^j\; e^{-{\cal S}_{mm}}
\right]=0~~.
\end{equation}
Using:
\begin{equation}
\frac{\partial}{\partial \hp_i\,^j}
\left(\frac{1}{z-\hp}\right)_i\,^j = \left(\tr \frac{1}{z-\hp}\right)^2~~,
\end{equation}
this leads to:
\begin{equation}
\label{omloop0}
\left\langle \om(z)^2 -\frac{g}{\hn} \tr \left( \frac{W'(\hp)}{z-\hp}
\right) - k(z) \right\rangle =0~~.
\end{equation}
Equation (\ref{tilde_f}) allows us to write (\ref{omloop0}) in the
form (\ref{loop.om}).

In order to find an additional equation for $k(z)$, we consider the
identity:
\begin{equation}
\int d\hp\, d\ha\, d\hs\, d {\hat Q}~
\frac{\partial}{\partial \ha_{ij}} \left[
\left(\frac{1}{z-\hp}\ha \frac{1}{z-\hp^T}\right)_{ij}\; e^{-{\cal S}_{mm}}
\right]=0~~.
\end{equation}
This implies equation (\ref{loop.m}) upon using the relations:
\begin{equation}
\label{identity2}
\tr \left(\hs \frac{\hp}{z-\hp} \ha \frac{1}{z-\hp^T}\right)=
-\tr \left(\hs \frac{1}{z-\hp} \ha \frac{\hp^T}{z-\hp^T}\right) =
\tr \left( \hs \frac{1}{z-\hp}\ha\right)\,,
\end{equation}
which follow trivially from the symmetry properties of ${\hat S}$ and
${\hat A}$ and invariance of the trace under
transposition\footnote{For this, it is useful to notice that
  $\tr({\hat S}\frac{1}{z-{\hat \Phi}}{\hat A}\frac{1}{z-{\hat
      \Phi}^T})=0$.}.

We next consider the identity:
\begin{equation}
\int d\hp\, d\ha\, d\hs\, d {\hat Q}~ 
\frac{\partial}{\partial \hs^{ij}} \left[
\left(\frac{1}{z-\hp^T}\hs \frac{1}{z-\hp}\right)^{ij}\; e^{-{\cal S}_{mm}}
\right]=0~~,
\end{equation}
which gives:
\begin{equation}
\left\langle
\frac{1}{2}\om(z)^2 - \frac{1}{2}\frac{g}{\hn} \om'(z) - k(z)
-\frac{g}{\hn} m_Q(z) \right\rangle =0~~.
\end{equation}
Together with (\ref{loop.m}), this implies the third loop equation
(\ref{loop.mq}).

Finally, we can derive a relation involving the flavors. For this, we start with the 
identity:
\begin{equation}
\label{Q_loop_mod}
\int d\hp\, d\ha\, d\hs\, d{\hat Q}~
\frac{\partial}{\partial \hat{Q}_{f}^i} \left[ 
\left(\frac{\lambda_{fg}}{z-\hp}\right)_{ij}\hat{Q}_{g}^j  \;
e^{-{\cal S}_{mm}}
\right]=0~~,
\end{equation}
where $\lambda$ is an arbitrary matrix in flavor space.

Equation (\ref{Q_loop_mod}) gives:
\begin{equation}
\label{intmd1_mod}
\left\langle \lambda_{ff}\omega(z) - 2 \sum_{g=1}^{{\hat N}_F} 
{\hat{Q}^T_f {\hat S} \frac{\lambda_{fg}}{z-\hp} \hat{Q}_g} \right\rangle =0~~.
\end{equation}
Since $\lambda$ is arbitrary, this implies:
\begin{equation}
\langle\omega(z)\delta_{fg}-2 \hat{Q}^T_f {\hat S} \frac{1}{z-\hp} \hat{Q}_g \rangle =0~~.
\end{equation}
Setting $f=g$ and summing over $f$ gives equation (\ref{loop.l}).

Since the matrix model admits the $O({\hat N}_F)$ flavor symmetry
(\ref{flavor_m}), it follows that the expectation value of any flavor
two-tensor must be $O({\hat N}_F)$ invariant and thus:
\begin{equation}
\langle \hat{Q}^T_f {\hat S} \frac{1}{z-\hp} \hat{Q}_g \rangle=\frac{1}{{\hat N}_F}\langle l(z)\rangle  \delta_{fg}~~.
\end{equation}
Combining with (\ref{loop.l}), this gives:
\begin{equation}
\langle \hat{Q}^T_f {\hat S} \frac{1}{z-\hp} \hat{Q}_g\rangle
=\frac{1}{2} \langle \omega(z) \rangle \delta_{fg}~~, 
\end{equation}
which is the matrix model analogue of equation (\ref{Qfg}).

\subsection{The eigenvalue representation}
\label{eigenvaluerep}

As mentioned above, the integration in (\ref{Z_mod}) should be performed
over an appropriate multidimensional contour inside the space of
complex matrices ${\hat \Phi}$, ${\hat S}$, ${\hat A}$ and ${\hat
  Q}_f$.  We shall specify this contour by first fixing the gauge
through diagonalizing \footnote{The fact that $\Phi$ is diagonalizable
  is part of the definition of our multidimensional contour. As in
  \cite{holo}, this implements a point-splitting regularization of the
  matrix integral, which can be removed trivially because the
  resulting Vandermonde determinant (to be described below) vanishes
  for coinciding eigenvalues.} ${\hat \Phi}$ and imposing conditions
on the remaining matrices {\em after} gauge-fixing. This procedure is
clearly gauge-invariant in the sense that it defines a
multidimensional contour which is stabilized by the action of the
complex gauge group $GL({\hat N},\C)$. Using the gauge symmetry to
bring ${\hat \Phi}$ to the form:
\begin{equation}
\label{gauge}
{\hat \Phi}={\rm diag}(\lambda_1\dots \lambda_\hn)~~
\end{equation}
allows us to write the matrix model action (\ref{action}) as:
\begin{equation}
{\cal S}_{mm}=\sum_{i=1}^\hn{W(\lambda_i)}+\sum_{i<j}\left[(\lambda_i-\lambda_j)
{\hat A}_{ij}+2
\sum_{f=1}^{{\hat N}_F}{{\hat Q}_f^i{\hat Q}_f^j}\right]{\hat S}_{ij}+\sum_{i=1}^\hn{\left[\sum_{f=1}^{{\hat N}_F}{({\hat Q}_f^i)^2}\right]{\hat S}_{ii}}~~.
\end{equation}
To make sense of the remaining integral, we impose the conditions:
\begin{equation}
\label{reality}
\lambda_i\in \gamma~~,~~{\hat S}_{ij}\in i\R~~,~~{\hat A}_{ij}\in \R~~,~~{\hat Q}_f^i\in \R~~,
\end{equation}
where $\gamma$ is an open contour in the complex plane whose
asymptotic behavior is dictated by the highest degree term in $W$ as
explained in \cite{holo} (one can take $\gamma$ to coincide with the
real axis if and only if $W$ is a polynomial of even degree). This
amounts to choosing:
\begin{eqnarray}
\Gamma&=&\{({\hat \Phi}, {\hat S}, {\hat A}, {\hat Q}_1\dots {\hat
  Q}_{{\hat N}_F})\in {\cal M}| ~\exists U \in GL({\hat N},\C){\rm ~such~that~}\nn \\
& &U\hp U^{-1}={\rm diag}(\lambda_1\dots \lambda_\hn) {\rm ~with~}
\lambda_1\dots \lambda_\hn \in \gamma {\rm~and~} \nn\\
& &U^{-T} {\hat S} U^{-1}\in Mat({\hat N},{\hat N},\R), ~U{\hat A}U^T\in Mat({\hat N},
{\hat N},\R), \nn\\
& &U{\hat Q}_f\in Mat({\hat N},1,\R) {\rm ~for~all~}f \}~~.
\end{eqnarray}

With this choice of integration contour, we obtain:
\begin{equation}
\label{ev}
Z=(2\pi i)^{\hn(\hn+1)/2} Z_Q Z_{red}~~,
\end{equation}
where:
\begin{equation}
\label{ZQ}
Z_Q=\int_{\R^{{\hat N}_F~{\hat N}}}{d\hq 
  \prod_{j=1}^{\hat N}{\delta(\sum_{f=1}^{{\hat N}_F}{(\hq_f^j)^2})}}~~
\end{equation}
and:
\begin{equation}
\label{Zred}
Z_{red}=\int_\gamma{d\lambda_1}\dots
\int_\gamma{d\lambda_\hn}{\prod_{i<j}{(\lambda_i-\lambda_j)}
e^{-\frac{\hn}{g}\sum_{i=1}^\hn{W(\lambda_i)}}}~~.
\end{equation}
To arrive at (\ref{ev}), we performed the integrals over 
${\hat S}_{ij}$, which appear linearly in the action.  This gives a
product of delta-functions which allows us to reduce the integral to
(\ref{ev}). From this expression, it is clear that the variables
${\hat A}$, ${\hat S}$ and ${\hat Q}_f$ decouple from ${\hat \Phi}$.
The interesting dynamics of the model is contained in the reduced
partition function (\ref{Zred}), which differs from that of a usual
(adjoint) one-matrix model only because the
Vandermonde determinant $\Delta=\prod_{i<j}{(\lambda_i-\lambda_j)}$ is
{\em not} squared in (\ref{Zred}). As we shall see below, $Z_{red}$
can in fact be identified with the partition function of a holomorphic
$SO({\hat N},\C)$ -- invariant one matrix model with a single
symmetric field ${\hat X}$ and action $\tr W({\hat X})$. This, of
course, is just the holomorphic matrix model associated with the
$SO(N)$ theory with symmetric matter via the Dijkgraaf-Vafa
conjecture.

\paragraph{Observation:}

The integral (\ref{ZQ}) is finite and non-vanishing precisely in the case of interest ${\hat N}_F=2$.
In this case, one easily checks that:
\be
\label{ZQ_ev}
Z_Q=\pi^{\hat N}~~.
\ee
For ${\hat N}_F>2$ we find $Z_Q=0$, while for 
${\hat N}_F=1$ we have $Z_Q=\infty$. This agrees with our previous discussion.

\subsection{Relation to the matrix model of the $SO(N)$ theory}
\label{relation}

As mentioned above, it turns out that the reduced model described by
(\ref{Zred}) agrees with the matrix model associated with the $SO(N)$
theory with symmetric matter via the Dijkgraaf-Vafa correspondence.
The Hermitian version of the latter is defined through:
\begin{equation}
\label{Zs}
Z_s:=\frac{1}{{|G_s|}}\int{d{\hat X} e^{-\frac{\hat N}{g}\tr W({\hat X})}}~~,
\end{equation}
where $|G_s|=\vol(SO({\hat N},\R)/S_{\hat N}) $ and ${\hat X}$ is a real
symmetric ${\hat N} \times {\hat N}$ matrix (thus ${\hat X}^T={\hat X}$ and
${\hat X}^\dagger={\hat X}$). In the Hermitian case, the measure is given by:
\begin{equation}
\label{measure_s}
d{\hat X}=\prod_{i\leq j}{d{\hat X}_{ij}}~~.
\end{equation}
The partition function (\ref{Zs}) and the measure (\ref{measure_s})
are invariant under the transformations:
\begin{equation}
\label{gauge_s}
{\hat X}\rightarrow V {\hat X} V^T~~,
\end{equation}
where $V$ is an element of $SO({\hat N},\R)$.

One way to see the aforementioned correspondence is by relating the
loop equations of the two models. The loop equation for the model
(\ref{Zs}) was extracted in \cite{KRS}, and involves only the
resolvent:
\begin{equation}
\omega_{s}(z) = \frac{g}{\hn} \tr \frac{1}{z-{\hat X}}~~.
\end{equation}
This loop equation takes the form:
\begin{equation}
\label{sloop}
\left\langle \om_{s} (z)^2 - \frac{g}{\hn} \om'_{s}(z) - 2 W'(z) \om_{s}(z)
+ 2 \tilde{f_{s}}(z) \right\rangle_s= 0~~,
\end{equation}
where $\langle \dots \rangle_s$ denotes averages computed in the model
(\ref{Zs}) and ${\tilde f}_{s}(z)$ is the random polynomial:
\begin{equation}
{\tilde f}_{s}(z):= \frac{g}{\hn} \tr\frac{W'(z)-W'({\hat X})}{z-{\hat X}}~~.
\end{equation}
Relation  (\ref{sloop}) corresponds to our equation (\ref{omega_loop}) under the
identifications:
\begin{eqnarray}
\label{matrix_ids}
\omega(z)&\longleftrightarrow& \omega_{s}(z)\\
{\tilde f}(z)&\longleftrightarrow& {\tilde f}_{s}(z)~~.
\end{eqnarray}
Since the quantities $k(z), m_Q(z), l(z)$ are determined by a solution
of (\ref{omega_loop}) via equations (\ref{loop_sol}), this gives a one
to one correspondence between solutions of the two models' loop
equations.

A more direct relation between the two models can be extracted from
their eigenvalue representations. To see this, we must first discuss
the eigenvalue representation of (\ref{Zs}).

In the context of the Dijkgraaf-Vafa correspondence, one must use the
holomorphic \cite{holo} version of (\ref{Zs}).  This is obtained by
allowing ${\hat X}$ to be a complex symmetric matrix (thus removing
the hermiticity constraint ${\hat X}^\dagger={\hat X}
\Longleftrightarrow {\bar {\hat X}}={\hat X}$) and considering gauge
transformations of the form (\ref{gauge_s}), where now $V$ is an
invertible complex-valued matrix subject to the constraints
$V^T=V^{-1}$ and $\det V =1$. This amounts to working with the
complexified gauge group $SO({\hat N},\C)$ (then the normalization
prefactor $|G_s|$ is also modified as explained in \cite{holo}). In
that case, (\ref{measure_s}) becomes the natural top holomorphic form
on the complex space ${\cal N}=\{{\hat X}\in Mat(N,\C)|{\hat
  X}^T={\hat X}\}$ and the integral in (\ref{Zs}) must be performed
along a real, boundary-less submanifold $\Delta$ of this space
whose dimension equals the complex dimension of ${\cal N}$ and whose closure is non-compact. 
The admissible choices of $\Delta$ are constrained by the requirement that
$\Delta$ be stabilized by the action (\ref{gauge_s}) of the
complexified gauge group and that the integral (\ref{Zs}) converge
when calculated along $\Delta$. This constrains the choice of $\Delta$
in terms of the leading coefficient of $W$ \cite{holo}.

To see the relation to (\ref{Zred}) explicitly, it suffices to notice
that integrating out the angular variables in (\ref{Zs}) leads to:
\begin{equation}
\label{Zs_ev}
Z_s=\int_\gamma{d\lambda_1}\dots
\int_\gamma{d\lambda_\hn}{\prod_{i<j}{(\lambda_i-\lambda_j)}
e^{-\frac{\hn}{g}\sum_{i=1}^\hn{W(\lambda_i)}}}~~,
\end{equation}
where $\gamma$ is a suitable open contour in the complex plane (since
$\gamma$ is constrained only by the leading term of $W$ \cite{holo},
it can be chosen to coincide with the contour used in (\ref{Zred})).
The form (\ref{Zs_ev}) corresponds to choosing $\Delta:=\{{\hat X}\in
{\cal N}|~\exists V\in SO({\hat N},\C) {\rm~such~that~} V{\hat X}
V^T={\rm diag}(\lambda_1\dots \lambda_\hn)~{\rm with}~\lambda_1\dots
\lambda_\hn\in \gamma\}$.  Note that the real dimension of $\Delta$
equals the complex dimension of ${\cal N}$, as required. 
Indeed, a generic complex symmetric matrix ${\hat X}$ can be diagonalized by a 
complex orthogonal transformation $V\in SO({\hat N},\C)$:
\begin{equation}
V{\hat X}V^T={\rm diag}(\lambda_1\dots\lambda_\hn)
\end{equation}
with complex $\lambda_j$. Since this is true outside a complex codimension one locus in ${\cal N}$, it 
follows that imposing the constraints $\lambda_j\in
\gamma$ simply halves the number of real parameters in ${\hat X}$.

The factor $\prod_{i<j}{(\lambda_i-\lambda_j)}$ is just the square
root of the factor $\prod_{i<j}{(\lambda_i-\lambda_j)^2}$ familiar
from the case of the adjoint representation of $GL(\hn, \C)$. This can be seen most
easily in the holomorphic matrix model set-up by repeating the
argument given in Appendix $A$ of \cite{holo} with the observation
that the number of integration variables in (\ref{measure_s}) is
reduced with respect to the case treated there due to the condition
$X_{ij}=X_{ji}$. Obviously (\ref{Zs_ev}) coincides with the reduced
matrix integral (\ref{Zred}):
\begin{equation}
Z_{red}=Z_s~~.
\end{equation}
Using relation (\ref{ev}), we find that the partition function of
(\ref{Z_mod}) can be written as:
\begin{equation}
\label{Z_fact}
Z=(2\pi i)^{\hn(\hn+1)/2} Z_s Z_{Q}~~,
\end{equation}
where $Z_Q$ is the quantity defined in equation (\ref{ZQ}). The
factorization (\ref{Z_fact}) shows that the average in our model of
any functional $F({\hat \Phi})$ which does not depend on $\hs$, $\ha$ or
$\hq_f$ coincides with the average of $F({\hat X})$ in the $SO({\hat
  N})$ model (\ref{Zs}):
\begin{equation}
\langle F({\hat \Phi})\rangle=\langle F({\hat X})\rangle_{s}~~.
\end{equation}
This explains the identifications (\ref{matrix_ids}) and gives the
precise relation between the two models.

\subsection{The resolvent loop equation from the eigenvalue representation}
\label{eigenvalue}

To derive the loop equation of (\ref{Zred}) (or, equivalently, of
(\ref{Zs})), one can follow standard procedure by starting with the
identity:
\begin{equation}
\int{\prod_{i=1}^\hn{d\lambda_i} \sum_{i=1}^\hn{\frac{\partial}{\partial
\lambda_i}
\left[\frac{1}{z-\lambda_i}\prod_{p<q}{(\lambda_p-\lambda_q)}e^{-\frac{\hn}{g}\sum_{r=1}^\hn{W(\lambda_r)}}\right]}}=0~~.
\end{equation}
Performing the partial derivative gives the relation:
\begin{equation}
\label{intmd}
\langle
\sum_{i=1}^\hn{\frac{1}{(z-\lambda_i)^2}}-\frac{\hn}{g}\sum_{i=1}^\hn{\frac{W'(\lambda_i)}{z-\lambda_i}}+
\sum_{i\neq j}{\frac{1}{\lambda_i-\lambda_j}\frac{1}{z-\lambda_i}}\rangle=0~~.
\end{equation}
Introducing the traced resolvent (\ref{omega}) and using the identity:
\begin{equation}
\sum_{i\neq j}{\frac{1}{\lambda_i-\lambda_j}\frac{1}{z-\lambda_i}}=
\frac{1}{2}\left[\sum_{i, j}{
\frac{1}{z-\lambda_j}\frac{1}{z-\lambda_i}}- \sum_{k} \frac{1}{(z-\lambda_k)^2}
\right]
\end{equation}
allows us to write (\ref{intmd}) in the form:
\begin{equation}
\label{loop1}
\langle \omega(z)^2-\frac{g}{\hn}\omega'(z)-\frac{2g}{\hn}
\sum_{i=1}^\hn{\frac{W'(\lambda_i)}{z-\lambda_i}}\rangle=0~~.
\end{equation}
Using the degree $d-1$ polynomial (\ref{tilde_f}):
\begin{equation}
\label{fdef}
\tilde f(z)=\frac{g}{\hn}\sum_{i=1}^\hn{\frac{W'(z)-W'(\lambda_i)}{z-\lambda_i}}~~
\end{equation}
leads to the loop equation in the form:
\begin{equation}
\label{loop2}
\langle \omega(z)^2-\frac{g}{\hn}\omega'(z)-2W'(z)\omega(z)+2\tilde f(z)\rangle=0~~,
\end{equation}
which recovers (\ref{omega_loop}).

Consider the spectral density
$\rho(\lambda)=\frac{1}{\hn}\sum_{i=1}^\hn{\delta(\lambda-\lambda_i)}$ and the expansions:
\bea
\langle \omega(z)\rangle &=&\sum_{j\geq 0}{\left(\frac{g}{\hn}\right)^j\omega_j(z)}\label{Nexpansion.omega}\\
\langle \rho(\lambda)\rangle &=&\sum_{j\geq 0}{\left(\frac{g}{\hn}\right)^j\rho_j(\lambda)}\\
\langle {\tilde f}(z)\rangle &=&\sum_{j\geq 0}{\left(\frac{g}{\hn}\right)^j{\tilde f}_j(z)}~~.
\eea
In the large $\hn$ limit, we have:
\begin{eqnarray}
\omega_0(z) &:=&\lim_{\hn\rightarrow \infty}\langle
\omega(z)\rangle=g\int{d\lambda \frac{\rho_0(\lambda)}{z-\lambda}}\nn\\
\tilde f_0(z)&:=&\lim_{\hn\rightarrow \infty}\langle \tilde f(z)\rangle= 
g\int{d\lambda
\rho_0(\lambda)\frac{W'(z)-W'(\lambda)}{z-\lambda }}~~.
\end{eqnarray}
In this limit, equation (\ref{loop1}) becomes:
\begin{equation}
\label{loop_planar}
\omega_0(z)^2-2W'(z)\omega_0(z)+2 {\tilde f}_0(z)=0~~,
\end{equation}
which shows that the quantity $u_0(z)=\omega_0(z)-W'(z)$ is a branch
of the hyperelliptic Riemann surface:
\begin{equation}
u^2=W'(z)^2-2\tilde f_0(z)~~.
\end{equation}

The loop equation (\ref{loop1}) can also be written in the form:
\begin{equation}
\label{loop3}
\langle \omega(z)^2-\frac{g}{\hn}\omega'(z) -2\oint_{{\cal C}}{\frac{dx}{2\pi i}
\frac{W'(x)\omega(x)}{z-x}}\rangle=0
\end{equation}
where ${\cal C}$ is a counterclockwise contour encircling all of the
eigenvalues $\lambda_i$ of ${\hat \Phi}$ but not the point $z$. The
form (\ref{loop3}) can be used to give an iterative solution for the
coefficients $\omega_j(z)$ of the large $\hn$ expansion
(\ref{Nexpansion.omega}).  In this paper, we are interested in the $\C\P^1$ 
and $\R\P^2$ diagram contributions $\omega_0(z)$  and  $\omega_1(z)$.
Expanding (\ref{loop3}) to order $O(g/\hn)$ and using the relation:
\begin{equation}
\langle \omega(z)^2\rangle=\langle \omega(z)\rangle
^2+O((g/\hn)^2)=\omega_0(z)^2+\frac{2g}{\hn}\omega_0(z)\omega_1(z)+ O((g/\hn)^2)~~,\label{expansion.omega.squared}
\end{equation}
we find:
\begin{eqnarray}
& &\omega_0(z)^2-2\oint_{\cal C}{\frac{dx}{2\pi
i}\frac{W'(x)\omega_0(x)}{z-x}}=0\label{loop_int0}\\ &
&2\omega_0(x)\omega_1(x)-2\oint_{\cal C}{\frac{dx}{2\pi
i}\frac{W'(x)\omega_1(x)}{z-x}}=\omega_0'(x)~~.\label{loop_int1}
\end{eqnarray}
The first relation is equivalent with the large $\hn$ limit
(\ref{loop_planar}) of the loop equation and determines $\omega_0(z)$
in terms of the polynomial ${\tilde f}_0(z)$. The second relation is
an inhomogeneous integral equation for $\omega_1(z)$, which constraints 
this quantity once ${\tilde f}_0(z)$ (and thus
$\omega_0(z)$) has been fixed.

\section{Relation between the matrix model and field theory}
\label{matrix.vs.field}
\subsection{Comparison of loop equations and Konishi constraints}
\label{konishi.vs.loop}

In this subsection we find a map from matrix
model to field theory quantities which takes the loop equations
(\ref{omega_loop}) and (\ref{loop_sol}) into the Konishi constraints
(\ref{konishi.R}), (\ref{konishi.T}) and (\ref{konishi_sol}).
We set $\hn_F=2$ and $N_F=8$ from now on. 

We start by considering a complex\footnote{Since our matrix model does not have a real counterpart, the 
set-up of \cite{holo} is essential in our case.} microcanonical
ensemble for our holomorphic matrix model following the procedure of \cite{holo}. This is obtained by
introducing complex chemical potentials $\mu_k$ for a partition of the complex plane
into domains $D_k$ with smooth boundary, followed by a Legendre transform
of the resulting grand-canonical generating function ${\cal
  F}(t,\mu)=-\left(\frac{\hat N}{g}\right)^2 \ln {\cal Z}(t, \mu)$, which replaces
the chemical potentials by complex variables $S_k$\footnote{These should
not be confused with the eigenvalues of the symmetric matrix S!} (remember that $t_j/j$ are the coefficients 
of $W$). This
produces the desired free energy (=microcanonical generating function)
$F(t, S)$, which satisfies the equations:
\begin{equation}
\frac{\partial F}{\partial S_k}=\mu_k~~.
\end{equation}
Working with this microcanonical ensemble amounts to imposing
the constraints:
\begin{equation}
\label{Sc_matrix}
\langle f_k \rangle =S_k
\end{equation}
where $f_k$ is the filling fraction of the domain $D_k$, which we
define through:
\begin{equation}
\label{eq:def.fk}
f_k=\oint_{\Gamma_k}{\frac{dz}{2\pi i}\omega(z)}~~.
\end{equation}
Here $\Gamma_k$ is the boundary of $D_k$. Note that the  variables $S_k$ must satisfy:
\be
\label{S_constraint}
\sum_{k}{S_k}=g~~,
\ee
so this a `constrained' microcanonical ensemble, as discussed in more detail in \cite{holo}. Thus passing to the 
microcanonical ensemble allows us to eliminate $g$ in terms of $S_i$, if we choose to treat $S_i$ as independent 
variables.

Expanding (\ref{Sc_matrix}) to $O(g/\hn)$ inclusively gives:
\begin{equation}
\label{Sc_m1}
\oint_{\Gamma_k}{\frac{dz}{2\pi i}\omega_0(z)}=S_k
\end{equation}
and:
\begin{equation}
\label{Sc_m2}
\oint_{\Gamma_k}{\frac{dz}{2\pi i}\omega_1(z)}=0~~.
\end{equation}

Imposing these conditions fixes all coefficients of ${\tilde f}_0(z)$ and
${\tilde f}_1(z)$. In fact, condition (\ref{Sc_m1}) singles out one solution $\omega_0$ of 
(\ref{loop_int0}) while (\ref{Sc_m2}) specifies the associated solution of (\ref{loop_int1})
by selecting the trivial solution of the associated linear homogeneous equation.
Then a simple counting argument along the lines of
\cite{us} shows that that $\langle \omega(z)\rangle$ is completely determined  as a function of $t_j$ and $S_k$. 
This also determines the expectation 
values of $k(z), m_Q(z)$ and $l(z)$ via equations (\ref{loop_sol}). 

Let us now consider the Konishi constraints (\ref{konishi.R}) and (\ref{konishi.T}) of the field theory, 
which determine all relevant quantities through equations (\ref{konishi_sol}). To specify uniquely a solution 
$(R,T)$ of (\ref{konishi.R}) and (\ref{konishi.T}), we shall impose the constraints:
\begin{equation}
\label{ft_constraints}
\oint_{\Gamma_k}{\frac{dz}{2\pi i}R(z)}=\cs_k~~,~~
\oint_{\Gamma_k}{\frac{dz}{2\pi i}T(z)}=N_k~~.
\end{equation}
Note that we must impose {\em two conditions} on
the solutions of (\ref{konishi.R}), (\ref{konishi.T}), which amounts
to fixing the coefficients of both the polynomials $f$ and $c$. As in \cite{cdsw}, the second equation in 
(\ref{ft_constraints}) can be viewed as a rigorous quantum definition of the rank of the $k$-th factor of the unbroken 
low energy gauge group. This interpretation requires that we choose $\Gamma_k$ such that they separate the critical points of 
$W$, which we shall assume from now on. 

To map a solution of the matrix loop equations to a solution of the Konishi constraints, we shall identify:
\be
\label{Sid}
\cs_k = S_k ~~.
\ee
Using (\ref{S_constraint}), we find that 
(\ref{Sid}) fixes the value of the matrix coupling constant in terms of field theory data:
\be
\label{g}
g=\sum_{k}{\cs_k}
\ee

Consider the large ${\hat N}$ expansions:
\begin{eqnarray}
\langle \omega(z)\rangle
&=&\sum_{j=0}^\infty{\left(\frac{g}{\hn}\right)^j\omega_j(z)}\,,
\label{expansion.omega}\\
\label{expansion.mq}
\langle k(z)\rangle &=& \sum_{j=0}^\infty \left(\frac{g}{\hat N}\right)^j m_j(z) ,\label{expansion.m}\\
\langle m_Q(z) \rangle &=&  \sum_{j=0}^\infty \left(\frac{g}{\hat N}\right)^j m_Q{}_j(z) \,,\\
\langle l(z) \rangle &=&  \sum_{j=0}^\infty \left(\frac{g}{\hat N}\right)^j l{}_j(z) \,.
\end{eqnarray}
Expanding (\ref{omega_loop}) to leading order, we find:
\begin{equation}
\label{omega_loop0}
\omega_0(z)^2-2W'(z)\omega_0(z)+2{\tilde f}_0(z)=0~~.
\end{equation}

We also expand (\ref{loop_sol}) to leading order and order $g/{\hat
  N}$ to obtain:
\begin{eqnarray}
\label{loop_sol0}
 k_0(z)&=&\frac{1}{2}\omega_0(z)^2\nn\\
m_{Q0}(z) &=&-\omega_0'(z) \\
l_0(z)&=& \omega_0(z)
\nn~~.
\end{eqnarray}
Comparing (\ref{omega_loop0}) with (\ref{konishi.R}) shows that
$\omega_0(z)$ and ${\tilde f}_0(z)$ should be identified with $R(z)$
and $f(z)$ respectively. Moreover, equations (\ref{loop_sol0}) agree
with the first and the last two equations in (\ref{konishi_sol})
provided that we identify $K(z)$ with $k_0(z)$ as well as $M_Q(z)$
with $4 m_{Q0}(z)$ and $L(z)$ with $4 l_0(z)$.

To recover (\ref{konishi.T}) and the second equation in
(\ref{konishi_sol}), we consider the $g/{\hat N}$ terms of
(\ref{omega_loop}) and of the first equation in (\ref{loop_sol}),
which read:
\begin{equation}
\label{omega_loop1}
2\omega_0(z)\omega_1(z)-\omega_0'(z)-2W'(z)\omega_1(z)+2{\tilde f}_1(z)=0
\end{equation}
and:
\begin{equation}
\label{loop_sol1}
 k_1(z)=\omega_0(z)\omega_1(z)+\frac{1}{2}\omega_0'(z)~~.
\end{equation}
Consider the operator: 
\be
\label{delta}
\delta:=\sum_{k}{N_k\frac{\partial}{\partial S_k}}~~.
\ee
Applying this to both sides of relation (\ref{omega_loop0}) and of the first equation in
(\ref{loop_sol0}) gives:
\begin{equation}
\label{omega_loopg}
2\omega_0(z)\delta \omega_0(z)-2W'(z)\delta \omega_0(z)+2\delta {\tilde f}_0(z)=0
\end{equation}
and:
\begin{equation}
\label{loop_solg}
 \delta k_0(z)=\omega_0(z)\delta \omega_0(z)~~.
\end{equation}
Combining these two equations with (\ref{omega_loop1}) and
(\ref{loop_sol1}) leads to the relations:
\begin{equation}
\label{omega_loop_k}
\omega_0(z)[\delta \omega_0(z)+4\omega_1(z)]
-W'(z)[\delta \omega_0(z)+4\omega_1(z)]-2\omega_0'(z)+[\delta {\tilde f}_0(z)+4{\tilde f}_1(z)]=0
\end{equation}
and:
\begin{equation}
\label{loop_solk}
 \delta k_0(z)+4k_1(z)=2\omega_0'(z)+
\omega_0(z)[\delta \omega_0(z)+4\omega_1(z)]~~.
\end{equation}
These relations agree with (\ref{konishi.T}) and (\ref{konishi_sol})
provided that we identify $T(z)$ with
$\delta \omega_0(z)+4\omega_1(z)$ as well as $M(z)$ with
$\delta k_0(z)+4k_1(z)$ and $c(z)$ with $\delta {\tilde
    f}_0(z)+4{\tilde f}_1(z)$.

In conclusion, matrix model and field theory quantities must be
identified according to the table:
\begin{center}
\begin{tabular}{|c|c|}
\hline
Matrix~Model & Field Theory\\ \hline
$S_k$ & $\cs_k$ \\
$\omega_0(z)$ & $R(z)$ \\
$\delta \omega_0(z) + 4 \omega_1(z)$ & $T(z)$\\
$\delta k_0(z) + 4 k_1(z)$ & $M(z)$ \\
$k_0(z)$ & $K(z)$\\
$4m_{Q0}(z)$ & $M_Q(z)$\\
$4 l_0(z)$ & $L(z)$\\
$\tilde f_0(z)$ & $f(z)$\\
$\delta \tilde f_0(z) + 4 \tilde f_1(z)$ & $c(z)$ \\
\hline
\end{tabular}
\end{center}

\begin{center}
  {\footnotesize Table 1: Identification between field theory and
    matrix model quantities.}
\end{center}

This correspondence recovers that used in \cite{KRS} upon applying the
field theory and matrix model relations (\ref{field_ids}) and
(\ref{matrix_ids}), which connect our model with the $SO(N)$ theory.

\subsection{The effective superpotential}
\label{effective}

The identifications of the previous subsection
allow us to determine the field theory effective superpotential up to
a term independent of the coefficients $t_j$ of $W$. This contribution can be
identified independently by using the Veneziano-Yankielowicz
computation of Section \ref{VY_section}.

For this, we note the relation $\langle \tr \Phi^j\rangle=j\frac{\partial
  F}{\partial t_j}$, which implies:
\begin{equation}
\langle \omega(z)\rangle=\frac{d}{dW(z)} F
\end{equation}
where $\frac{d}{dW (z)}:=\sum_{j\geq
  0}{\frac{j}{z^{j+1}}\frac{\partial}{\partial t_j}}$.  Combing this
with the expansion:
\begin{equation}
F=\sum_{j\geq 0}{\left( \frac{g}{\hat N}\right)^j F_j}~~
\end{equation}
gives:
\begin{equation}
\omega_0(z)=\frac{d}{dW(z)} F_0~~,~~\omega_1(z)=\frac{d}{dW(z)} F_1~~.
\end{equation}
On the other hand, one has the obvious field theory relation:
\begin{equation}
\frac{\partial W_{eff}}{\partial t_j}=\frac{1}{j}\langle \tr \Phi^j\rangle 
\end{equation}
which gives:
\begin{equation}
T(z)=\frac{d}{d W(z)} W_{eff}~~.
\end{equation}

Using the identification $T(z)=\delta \omega_0(z)+4\omega_1(z)$ (where $\delta$ is the operator given in (\ref{delta})),
this implies:
\begin{equation}
\label{matrix.model.superpot}
W_{eff}(t,\cs)=\left[\sum_{i}{N_i \frac{\partial F_0}{\partial S_i}}+4 F_1+\psi(S)\right]_{S_i=\cs_i}~~.
\end{equation}
Here $\psi$ is a function which depends only $\cs_i$ but
not on the coefficients of $W$\footnote{Remember that relation (\ref{g}) eliminates $g$ 
in terms of $\cs_i$.}. Since we always set $S_i=\cs_i$, we shall only use the notation $\cs_i$  from now on.
Notice that we have derived (\ref{matrix.model.superpot}) without having to postulate 
some analytic continuation which would avoid identifying integer or real quantities with complex numbers. This is because the 
use the formalism of \cite{holo},  which automatically avoids such problems. 
Also notice the prefactor of $4$ in front of the $\R\P^2$ contribution $F_1$, which arises naturally in our derivation. 
The fact that diagrams of topology $\R\P^2$ generally
contribute with a factor $4$ to the effective superpotential was previously discussed in \cite{ino} by using the perturbative
superfield approach of \cite{dglvz}.

Expression (\ref{matrix.model.superpot}) determines the 
effective superpotential only up to the coupling-independent term $\psi(\cs)$. 
Together with the contribution to (\ref{matrix.model.superpot}) from the non-perturbative 
part of $F$, this term should correspond to the Veneziano-Yankielowicz 
potential computed in Subsection 2.3, which cannot be determined 
through Konishi anomaly arguments.
Applying the conjecture of \cite{DV,Ooguri} to our model leads to the proposal:
\be
\label{psi}
\psi(\cs)=\alpha \sum_{i=1}^{d}{\cs_i}~~,
\ee
where:
\be
\label{alpha}
\alpha=(N-4)\ln \Lambda~~
\ee
with $\Lambda$ the field theory scale of Subsection 2.3. 
To check this expression, we now proceed to compute the non-perturbative 
contribution  to $W_{eff}$
in the matrix model and compare with the results of Subsection 2.3. This will allow us to complete the proof
of the relation:
\be
\label{W_eff_final}
W_{eff}=\sum_{i}{N_i \frac{\partial F_0}{\partial \cs_i}}+4 F_1+\alpha \sum_{i=1}^{d}{\cs_i}~~.
\ee

\subsection{Computation of the Veneziano-Yankielowicz superpotential from the matrix model}
\label{VY.from.matrix}

In this subsection we show how the Veneziano-Yankielowicz contribution to the
effective superpotential can be extracted from the matrix model. We shall follow the approach of 
\cite{Ooguri,cdsw}, by computing the non-perturbative contribution to the matrix 
integral and checking agreement between the 
non-perturbative part of (\ref{W_eff_final}) and the result (\ref{VYpot}) of Subsection 2.3.

Let us  consider the classical matrix vacuum:
\be
\label{m_vac}
\langle\hp\rangle=\mathrm{diag}(\lambda_1 {\bf 1}_{{\hat N}_1}, \dots,
\lambda_d {\bf 1}_{{\hat N}_d})~~,~~\langle {\hat S}\rangle =\langle {\hat A}\rangle =\langle {\hat Q}_f \rangle=0~~.
\ee 
where $\lambda_j$ are the critical points of $W$. Following \cite{Ooguri, cdsw}, we shall compute the Gaussian approximation to the 
matrix integral expanded around this vacuum. This is the semiclassical approximation in the background (\ref{m_vac}). 

Since we wish to compare with field theory, we must work in the 
microcanonical ensemble, which constraints $g$ through relation (\ref{g}). In the semiclassical approximation about the background
(\ref{m_vac}), one has $\cs_i=\hn_i$ by equations (\ref{Sc_matrix}) and (\ref{eq:def.fk}). Then (\ref{g}) implies $g=\hn$, 
which means that the prefactor of the action in the exponential of the matrix integrand must be set to one. 
In particular, the large $\hn$ expansion can be reorganized in powers of 
$1/\hn$ rather than $g/\hn$. For simplicity, we can therefore start with:
\begin{equation}
  \label{eq:normalization}
  Z= e^{-F} = \frac{1}{\mathrm{vol}(G)} \int d\hp d\ha d\hs d\hq
e^{-\tr[ W(\hp) + \hs \hp \ha] - \sum_{f=1}^{2}{\hq_f^T \hs \hq_f}}
\end{equation}
and impose the microcanonical ensemble conditions $\cs_i=N_i$ after performing the semiclassical approximation.
We have set ${\hat N}_F=2$, since this is the only case when the matrix model is well-defined.

As in Subsections 4.3 and 4.4, the partition function (\ref{eq:normalization}) can be reduced to:
\begin{equation}
  \label{eq:zreduced}
  Z = (2\pi i)^{\hn(\hn+1)/2} \pi^\hn Z_s~~,
\end{equation}
where $Z_s$ is the partition function of the matrix model with $SO(\hn,\C)$ 
gauge group and a complex matrix ${\hat X}$ in the symmetric representation.

It is clear from (\ref{eq:zreduced}) that we can expand $Z$ around this vacuum 
by expanding $Z_s$ around the background $\langle {\hat X} \rangle =  \langle\hp\rangle$. 
We let $x:={\hat X}-\langle {\hat X} \rangle$ denote the fluctuations of ${\hat X}$ and
decompose $x$ into  $N_i\times N_j$ blocks $x_{ij}$. In the
reduced model $Z_s$, we have an $SO(\hn,\C)$ gauge symmetry which allows
us to set the off-diagonal blocks of $x$ to zero. Thus we can choose the gauge:
\be
x_{ij} = 0~~{\rm for}~~i\neq j~~.
\ee  
To implement this in the BRST formalism, we 
introduce  ghosts $C_{ij}$,
antighosts $\bar C_{ij}$ and Lagrange multipliers $B_{ij}$. 
These transform in the adjoint representation of $SO(\hn,\C)$, so 
$C_{ij}^T = - C_{ji}$,  $\bar C_{ij}^T = -\bar C_{ji}$ and
  $B_{ij}^T = - B_{ji}$.
The quadratic part of the gauge-fixing action is:
\begin{equation}
  \label{eq:gaugefixing}
  S_{g.f.} = \sum_{i< j}^d\tr ( B_{ij} x_{ij}  + i (\lambda_i-\lambda_j)
\bar C_{ij} C_{ij} )~~,
\end{equation}
where we used the BRST transformations:
\begin{equation}
  \label{eq:brst}
  s {\hat X} = i [C,X],~~~~ sC = i C^2,~~~~ s\bar C = B,~~~ sB =0~~.
\end{equation}  
When expanding to second order in $x$, the diagonal blocks $x_i:=x_{ii}$ acquire masses $m_i = W''(\lambda_i)$. 
Hence the non-perturbative piece of the partition function (\ref{eq:normalization}) is given by:
\begin{equation}
  \label{eq:zmore}
  Z_{np} = \frac{(2\pi i)^{\hn(\hn+1)} \pi^\hn}{\mathrm{Vol}(G')} 
\int \prod_{i=1}^d dx_{ii}
e^{-\frac{W''(\lambda_i)}{2} \tr( x_i ^2) }   
\int  \prod_{j< k} dB_{jk}  dC_{jk}  d\bar C_{jk}
e^{-S_{g.f.}}\,.
\end{equation}
Here $G'=\prod_{i=1}^d  SO(\hn_i)$ is the unbroken gauge group.
Since the action is quadratic, we can choose the integration contour: 
\begin{equation}
  \label{eq:contour}
  B_{ij} \in i \R\ ~~~~~ x_{ii} \in \R \,,
\end{equation}
which gives:
\begin{eqnarray}
  \label{eq:zfinal}
  Z_{np} =
\prod_{i<j}^d [(2 \pi)^{\hn_i \hn_j} (\lambda_i-\lambda_j)^{\hn_i \hn_j} ]
\prod_{i=1}^d \left[
\frac{(2\pi i)^{\frac{\hn_i(\hn_i+1)}{2}} \pi^{\hn_i}}{\mathrm{vol}(SO(\hn_i))}
\left(\frac{\pi }{ W''(\lambda_i)}\right)^{\frac{\hn_i(\hn_i+1)}{4}}\right]~~.
\end{eqnarray}
Using \cite{Roemelsberger}:
\begin{equation}
  \label{eq:vol.son}
 \log[ \mathrm{vol}(SO(\hn))]  = -\frac{\hn^2}{4}
\log\frac{\hn}{2\pi e^{3/2}}+ 
\frac{\hn}{4}\log\frac{2 \hn}{\pi e} + O(\hn^0)~~,  
\end{equation}
we extract the free energy $F_{np}=-\log Z_{np} $:
\begin{eqnarray}
  \label{eq:logZ}
  F_{np} &=& 
\sum_{i=1}^d \left[-\left(\frac{\hn_i^2}{4} +\frac{\hn_i}{4}\right) 
\log\frac{(-4\pi^3)}{W''(\lambda_i)}-
\frac{\hn_i^2}{4}  \log\frac{\hn_i}{2\pi e^{3/2}}
+\frac{\hn_i}{4}  \log\frac{2\hn_i}{\pi^5 e}\right]-
\nonumber\\
& & -\sum_{i < j} \hn_i \hn_j \log\left[2\pi (\lambda_i-\lambda_j)\right]~~.
\end{eqnarray}
As explained above, we have $\cs_i=\hn_i$ since we are in the semiclassical approximation of the matrix model. 
Replacing ${\hat N}_i$ by $\cs_i$ in (\ref{eq:logZ}), we can identify the quadratic terms in $\cs_i$
with $F_0^{np}$ and the linear terms with $F_1^{np}$:
\begin{eqnarray}
  \label{eq:f0f1}
  F_0^{np} &=& -\sum_{i=1}^d \frac{\cs_i^2}{4}
\log\left[ \frac{(-2\pi^2)\cs_i}{e^{3/2} W''(\lambda_i)}\right] - 
\sum_{i < j} \cs_i \cs_j \log\left[2 \pi (\lambda_i-\lambda_j)\right]\,,\\
F_1^{np} &=& -\sum_{i=1}^d \frac{\cs_i}{4} 
\log\left[  \frac{(-2 \pi^8 e)}{\cs_i W''(\lambda_i)} \right]~~.
\end{eqnarray}
According to (\ref{W_eff_final}), the non-perturbative contribution to the 
effective superpotential should be given by:
\begin{equation}
  \label{eq:weffchiral}
  W_{eff}^{np} = \sum_{i=1}^d N_i \frac{\partial F^{np}_0}{\partial \cs_i} + 4 F^{np}_1 + \alpha\sum_{i=1}^d{\cs_i}~~,
\end{equation}
Ignoring constant and linear terms in $\cs_i$ 
(which can be absorbed by a finite
renormalization of the field theory scale $\Lambda$ of Subsection
2.3), we find:
\begin{equation}
  \label{eq:finalsuperpot}
  W_{eff}^{np} = \sum_{i=1}^d \frac{\cs_i}{2} \log\left[ 
\frac{ {\displaystyle\Lambda^{2N-8} W''(\lambda_i)^{N_i+2} \prod_{j\neq i} (\lambda_i-\lambda_j)^{-2N_j}}}
{\cs_i^{N_i-2}}
\right] +O(\cs)~~,
\end{equation}
which recovers the leading contribution (\ref{VYpot}) derived by
threshold matching.

\section{Conclusions}
\label{conclusion}

We studied a class of non-anomalous, chiral ${\cal N}=1$ $U(N)$ gauge
theories with antisymmetric, conjugate symmetric, adjoint and
fundamental matter in the context of the Dijkgraaf-Vafa
correspondence.  By using the method of generalized Konishi anomalies,
we extracted a set of chiral ring constraints which allowed us to
identify the `dual' holomorphic matrix model and give an explicit expression for
the gaugino superpotential in terms of matrix model data. This
gives a proof of the Dijkgraaf-Vafa correspondence for a class of
theories with quite nontrivial chiral matter content.

As a by-product of this analysis, we found that the effective
superpotential of our models coincides with that produced upon
confinement in non-chiral $SO(N)$ field theories with a single chiral
superfield transforming in the symmetric representation.  This
provides an independent proof of a relation suggested by holomorphy
arguments.

Our results encourage us to think that similar methods could be
applied successfully in order to extract non-perturbative information
about more realistic models employed in supersymmetric phenomenology.
It would be interesting to see how far this program can be 
implemented.

Surprisingly, we found that the number of fundamental flavors $N_F$
in field theory must be taken to differ from the number of flavors ${\hat N}_F$ in
 the dual matrix model. This is unlike the non-chiral case considered in
\cite{argurio,rr,Radu_fundamentals1, Hofman,Radu_fundamentals2, OT2, Seiberg},
for which the number of fundamental flavors agrees between field theory and the matrix model.  
More precisely, consistency of the matrix model requires ${\hat N}_F=2$, while 
anomaly cancellation in our field theory requires $N_F=8$. 
Despite this disagreement, one can match the 
Konishi constraints in the chiral ring with the loop equations of the two-flavor 
matrix model. 

Another interesting question concerns the geometric engineering of our
models, which provides their embedding in IIB string theory and leads
to an interpretation of the gaugino superpotential as the
flux-orientifold superpotential of certain non-compact Calabi-Yau
backgrounds. As in the case of the $U(N)$ theory with symmetric or
antisymmetric matter \cite{us, Naculich, Naculich2} (whose geometric
engineering was discussed in \cite{llt}), this can be achieved by
applying the methods of \cite{OT}. A deeper understanding of the 
geometric realization could also shed light on the mismatch between the number of flavors in the matrix model and field theory. 
A detailed investigation of these
issues is under way \cite{chiral2}.

\acknowledgments{ This work was supported by DFG grant KL1070/2-1.  R.
  T. is supported by DOE Contract DE-AC03-76SF00098 and NSF grant
  PHY-0098840. K.L. would like to thank Andi Brandhuber for helpful
  discussions. C.I.L thanks the Kavli Institute for Theoretical Physics for hospitality and providing excellent 
conditions. }

\appendix

\section{Gauge invariance of the matrix model measure}
\label{measure_inv}

With an arbitrary number of flavors 
${\hat N}_F$, we have the following transformations under the action (\ref{matrix_gauge}):
\bea
\label{measure_gauge}
d\hp &\rightarrow& d\hp\nn\\
d\ha &\rightarrow& (\det U)^{\hn-1} d\ha\nn \\
d \hs &\rightarrow& (\det U)^{-(\hn+1)}  d\hs \\
d\hq &\rightarrow & (\det U)^{{\hat N}_F} d\hq~~,\nn
\eea
where $U$ is an arbitrary element of $GL({\hat N},\C)$. Therefore, the matrix model measure $d\mu=d\hp d\ha d\hs d\hq$ 
transforms as:
\be
d\mu \rightarrow (\det U)^{{\hat N}_F-2} d\mu~~,
\ee
and is invariant if and only if ${\hat N}_F=2$. 

The first relation in (\ref{measure_gauge}) is familiar from the (adjoint) holomorphic one-matrix model \cite{holo}, 
while the last relation is obvious. Thus we only have to prove the second and third equation.
For this, it suffices to check them for diagonalizable $U$, 
since our matter representation is continuous and the set of diagonalizable elements is dense in $GL({\hat N},\C)$.
Thus we can take $U=V T V^{-1}$, where $V$ is an element of $GL({\hat N},\C)$ and 
$T={\rm diag}(t_1\dots t_{\hat N})$, with $\prod_{j}{t_j}\neq 0$. The explicit form of 
the transformations (\ref{matrix_gauge}) gives:
\bea
d\ha &\rightarrow& [\det ~R_a(U)]~ d\ha\nn\\
d\hs &\rightarrow& [\det ~R_{cs}(U)]~ d\hp~~,
\eea
where $R_a$ and $R_{cs}$ are the antisymmetric and conjugate symmetric representations of $GL({\hat N},\C)$:
\bea
R_a(U)(A)&=&UAU^T~~,~~A^T=-A\nn\\
R_{cs}(U)(S)&=&U^{-T}S U^{-1}=U^{-T} S (U^{-T})^T~~,~~S^T=S~~. 
\eea
Using $R_a(VTV^{-1})=R_a(V)R_a(T)R_a(V)^{-1}$ and $R_{cs}(VTV^{-1})=R_{cs}(V)R_{cs}(T)R_{cs}(V)^{-1}$, we find:
\bea
d\ha &\rightarrow& [\det ~R_a(T)]~ d\ha\nn\\
d\hs &\rightarrow& [\det ~R_{cs}(T)]~ d\hs~~.
\eea
Since $T$ is diagonal, one easily obtains:
\bea
\det ~R_a(T)&=&\prod_{1\leq i<j\leq \hn }{t_it_j}=(\prod_i{t_i})^{\hn-1}\nn\\
\det ~R_{cs}(T)&=&\prod_{1\leq i\leq j\leq \hn }{t_i^{-1}t_j^{-1}}=(\prod_i{t_i})^{-(\hn+1)}~~,
\eea
so that:
\bea
\det ~R_a(U)=(\det U)^{\hn -1}~~{\rm and}~~\det ~R_{cs}(U)=(\det U)^{-\hn -1}~~.
\eea
This leads immediately to the second and third equation in (\ref{measure_gauge}).

\section{Classical vacua of the $SO(N)$ model}
\label{SO_moduli}

For the $SO(N)$ model, the only matter is the complex superfield $X$,
subject to the condition $X^T=X$ and transforming as follows under the
gauge group:
\begin{equation}
\label{Ogauge}
X\rightarrow V X V^T=VXV^{-1}~~.
\end{equation}
Here $V$ is valued in $SO(N,\R)$.  The F-flatness relations for $W_{tree}=\tr W(X)$ read:
\begin{equation}
\label{FO}
W'(X)=0
\end{equation}
while the D-flatness condition has the form:
\begin{equation}
\frac{1}{2i}[X,{\bar X}]=0~~,
\end{equation}
where the left hand side is the moment map\footnote{As usual, we used
  dualization with respect to the Killing form
  $(\xi,\eta)=-\tr(\xi\eta)$ of $o(N)$ in order to view the moment map
  as being valued in the Lie algebra $o(N)$ rather than in its dual.}
for the action (\ref{Ogauge}) of $SO(N,\R)$ (computed with respect to
the natural symplectic form $\omega(X,Y)=-\tr({\bar X}Y)$ on the
representation space).

Writing $X=X^{re}+iX^{im}$, where $X^{re}$ and $X^{im}$ are two
real-valued symmetric matrices, the D-flatness constraint becomes:
\begin{equation}
[X^{re}, X^{im}]=0~~,
\end{equation}
which shows that $X^{re}$ and $X^{im}$ can be diagonalized
simultaneously by performing a gauge transformation (\ref{Ogauge}):
\begin{eqnarray}
\label{Xdiag}
X^{re}&=&{\rm diag}(\lambda^{re}_1 {\bf 1}_{N_1}\dots 
\lambda^{re}_D {\bf 1}_{N_D})\nn\\
X^{im}&=&{\rm diag}(\lambda^{im}_1 {\bf 1}_{N_1}\dots \lambda^{im}_D {\bf 1}_{N_D})~~.
\end{eqnarray}
Here the real numbers $\lambda^{re}_k$ and $\lambda^{im}_k$ are such
that the pairs $(\lambda^{re}_k, \lambda^{im}_k)$ are mutually
distinct.

Writing $\lambda_k:=\lambda^{re}_k+i\lambda^{im}_k$, equations
(\ref{X_diag}) show that $X$ can be diagonalized via a gauge
transformation (\ref{Ogauge}):
\begin{equation}
\label{X_diag}
X={\rm diag}(\lambda_1 {\bf 1}_{N_1}\dots \lambda_D {\bf 1}_{N_D})~~,
\end{equation}
where $\lambda_k$ are distinct complex numbers.

The F-flatness condition (\ref{FO}) now shows that we can take
$D=d=\deg W'(z)$ and $\lambda_1\dots \lambda_d$ must coincide with the
distinct critical points of $W$ (again we use the convention that the
block $\lambda_k 1_{N_k}$ does not appear in (\ref{X_diag}) if
$N_k=0$). In such a vacuum, the gauge group is broken down to
$[\prod_{k=1}^d{O(N_k)}]/\{-1,1\}$.

\section{Derivation of the generalized Konishi constraints}
\label{konishi_proofs}

Let us outline the derivation of the generalized Konishi relations.
We will make heavy use of the chiral ring relations
(\ref{cr_w_alpha}) and (\ref{cr_phi})-(\ref{cr_q}).

The first two equations ({\ref{konishi.phi.one}-\ref{konishi.phi.two})
  are straightforward modifications of the relations found in
  \cite{cdsw} for the $U(N)$ model with a single adjoint field $\Phi$
  and no extra matter. The only differences are due to the coupling of
  $\Phi$ to the symmetric and antisymmetric fields and appear on the
  left hand sides of these relations.
  
  The derivation of the other relations is somewhat more involved. Let
  us consider first the symmetric field $S$ and the general
  transformation:
\begin{equation}
\delta S = X^T S Y
\end{equation}
where $X$ and $Y$ are arbitrary Grassmann even or odd $N\times N$
matrices. We start by computing the anomaly term induced by this
transformation.  For this, recall that ${\cal W}_\alpha$ acts on
$\delta S$ according to the conjugate symmetric representation:
\begin{equation}
{\cal W}_\alpha . \delta S = - {\cal W}_\alpha^T  \delta S - 
\delta S  {\cal W}_\alpha~~,
\end{equation}
where juxtaposition on the right hand side stands for matrix
multiplication.

Applying this formula twice, we find the anomaly term:
\begin{eqnarray}
-32 \pi^2 {\cal A} &:=&
\Tr\left({\cal W}^\alpha . {\cal W}_\alpha . \frac{\partial \delta S}{\partial S}\right) =\nn\\
&=&\frac{\partial (\delta S^{ir})}{\partial S^{ij}}{\cal W}^\alpha{}_r\,^s
{\cal W}_\alpha{} _s\,^j  + 
2 
{\cal W}^\alpha{} _r\,^i \frac{\partial (\delta S^{rs})}{\partial S^{ij}}
{\cal W}_{\alpha,s}\,^j 
+\,{\cal W}^\alpha{}_r\,^i {\cal W}_{\alpha,m}\,^r
\frac{\partial (\delta S^{mj})}{\partial S^{ij}}\,=\nn \\ 
&= & \frac{1}{2} \tr(X) \tr(Y{\cal W}^2) + \frac{1}{2} \tr(XY{\cal W}^2) + 
\tr({\cal W}^\alpha X) \tr(Y{\cal W}_\alpha) + \nn \\  
& &(-)^{[X]}\tr(X {\cal W}^\alpha Y{\cal W}_\alpha) + \frac{1}{2}
\tr({\cal W}^2 X) \tr(Y) + \frac{1}{2} \tr(X{\cal W}^2Y)~~.
\end{eqnarray}
where $[X]$ denotes the Grassmann parity of $X$, $\Tr(\dots)$ stands for
the trace in the conjugate symmetric representation and $\tr(\dots)$
is the trace in the fundamental representation.

Applying this relation for $X=Y=\frac{1}{z-\Phi}$, we obtain:
\begin{eqnarray}
\label{SXY.one.lhs}
{\cal A} &=&-\frac{1}{32\pi^2}
\left[\tr\frac{1}{z-\Phi}\tr \frac{{\cal W}^2}{z-\Phi} +
2 \tr  \frac{{\cal W}^2}{(z-\Phi)^2} +\tr \frac{{\cal W}^\alpha}{z-\Phi}
\tr \frac{{\cal W}_\alpha}{z-\Phi}\right] = \nn\\
&=&{\bf T}(z) {\bf R}(z) - 2 {\bf R}'(z) - \frac{1}{2} {\bf w}^\alpha {\bf w}_\alpha~~.
\end{eqnarray}

The left hand side of the generalized Konishi relation
(\ref{konishi_relation}) involves the variation of the superpotential
contracted with $\delta S$:
\begin{eqnarray}\label{SXY.one.rhs}
\delta S^{ij} \frac{\partial W}{\partial S^{ij}} &=&
-\frac{1}{2} \tr\left( \frac{1}{z-\Phi} A \frac{\Phi^T}{z-\Phi^T} S\right)+
\frac{1}{2} \tr\left( \frac{\Phi}{z-\Phi} A \frac{1}{z-\Phi^T} S\right) +\\
& & +Q_f^T\frac{1}{z-\Phi^T}S\frac{1}{z-\Phi}Q_f=
 \tr\left(S \frac{1}{z-\Phi}A\right) +Q_f^T\frac{1}{z-\Phi^T}S\frac{1}{z-\Phi}Q_f~~.\nn 
\end{eqnarray}
To arrive at this equation, we used the identity (\ref{identity2}).
Combining (\ref{SXY.one.rhs}) and (\ref{SXY.one.lhs}) leads to the
Konishi relation (\ref{konishi.S.one}).

We next consider the transformation:
\begin{equation}
\delta A = X A Y^T\,.
\end{equation}
Since:
\begin{equation}
{\cal W}_\alpha . \delta A = {\cal W}_\alpha \delta A + \delta A {\cal W}_\alpha^T~~,
\end{equation}
this leads to the anomaly term:
\begin{eqnarray}
-32 \pi^2 {\cal A} &:=&
\Tr\left({\cal W}^\alpha . {\cal W}_\alpha . \frac{\partial \delta A}{\partial A}\right) =\nonumber\\
&=&{\cal W}^\alpha{}_i\,^s
{\cal W}_\alpha{} _s\,^m
\frac{\partial (\delta A_{mj})}{\partial A_{ij}}  + 
2 
{\cal W}^\alpha{} _i\,^r \frac{\partial (\delta A_{rs})}{\partial A_{ij}}
{\cal W}_{\alpha,j}\,^s 
+\,\frac{\partial (\delta A_{ir})}{\partial A_{ij}}
{\cal W}^\alpha{}_s\,^r {\cal W}_{\alpha,j}\,^s
\,=\nonumber \\ 
&= & \frac{1}{2} \tr(X) \tr(Y{\cal W}^2) - \frac{1}{2} \tr(XY{\cal W}^2) + 
\tr({\cal W}^\alpha X) \tr(Y{\cal W}_\alpha) - \nonumber \\  
& &(-)^{[X]}\tr(X {\cal W}^\alpha Y{\cal W}_\alpha) + \frac{1}{2}
\tr({\cal W}^2 X) \tr(Y) - \frac{1}{2} \tr(X{\cal W}^2Y)\,.
\end{eqnarray}
Choosing $X=Y=\frac{1}{z-\Phi}$, one finds:
\begin{eqnarray}\label{AXY.one.lhs}
{\cal A} &=&-\frac{1}{32\pi^2}\left[\tr\frac{1}{z-\Phi} \tr \frac{{\cal W}^2}{z-\Phi} -
2 \tr  \frac{{\cal W}^2}{(z-\Phi)^2} + \tr\frac{{\cal W}^\alpha}{z-\Phi}
\tr\frac{{\cal W}_\alpha}{z-\Phi}\right] = \nn\\
&=& {\bf T}(z) {\bf R}(z) + 2 {\bf R}'(z) - \frac{1}{2} {\bf w}^\alpha 
{\bf w}_\alpha~~.
\end{eqnarray}
On the other hand, the choice $X=\frac{{\cal W}^\beta}{z-\Phi}$ and
$Y=\frac{{\cal W}_\beta}{z-\Phi}$ gives:
\begin{equation}\label{AXY.two.rhs} 
-32 \pi^2 {\cal A} =\tr \left(\frac{{\cal W}^\alpha{\cal W}^\beta
  }{z-\Phi}\right) \left(\frac{{\cal W}_\alpha{\cal W}_\beta
  }{z-\Phi}\right) = -(-32 \pi^2)^2\frac 1 2 {\bf R}(z)^2~~,
\end{equation}
where we used the spinor identities:
\begin{eqnarray}
\psi^\alpha \psi^\beta &=& \frac{1}{2} \epsilon^{\alpha\beta} \psi^2 \,,
\nonumber\\
\psi_\alpha \psi_\beta &=& -\frac{1}{2} \epsilon_{\alpha\beta} \psi^2 \,,
\nonumber\\
\epsilon_{\alpha\beta} \epsilon^{\beta\gamma} &=& \delta^\gamma_\alpha\,.
\end{eqnarray}

The left hand side of the anomaly equation for $X=Y=\frac{1}{z-\Phi}$
has the form:
\begin{equation}
\delta A_{ij} \frac{\partial W}{\partial A_{ij}} =   \tr\left(S \frac{1}{z-\Phi}A\right)\,,
\end{equation}
where again we used identity (\ref{identity2}). Together with
(\ref{AXY.one.lhs}). this leads to the Konishi relation
(\ref{konishi.A.one}).

If we chose $X=\frac{{\cal W}^\beta}{z-\Phi}$ and $Y=\frac{{\cal
    W}_\beta}{z-\Phi}$, the left hand side becomes:
\begin{eqnarray}
\delta A_{ij} \frac{\partial W}{\partial A_{ij}} &=& \frac 1 2 
\tr\left(S \frac{\Phi {\cal W}^\beta}{z-\Phi}A \frac{ {\cal W}_\beta^T}{z-\Phi^T}\right) -
\frac 1 2 
\tr\left(S \frac{{\cal W}^\beta}{z-\Phi}A \frac{ \Phi^T{\cal W}_\beta^T}{z-\Phi^T}\right)~~. 
\end{eqnarray} 
We next notice that:
\begin{equation}
\tr\left(S \frac{ {\cal W}^\beta}{z-\Phi}A \frac{ {\cal W}_\beta^T}{z-\Phi^T}\right) =
-\tr\left(\frac{ {\cal W}_\beta}{z-\Phi}A^T \frac{ ({\cal W}^\beta)^T}{z-\Phi^T} S^T\right)=
-\tr\left(S \frac{ {\cal W}^\beta}{z-\Phi}A \frac{ {\cal W}_\beta^T}{z-\Phi^T}\right)=0
\end{equation}
where in the first line we used invariance of the trace under
transposition and in the second line we used the symmetry of $S$ and
antisymmetry of $A$ together with the spinor identity
$\psi^\beta\psi_\beta =-\psi_\beta\psi^\beta$.  A similar derivation
shows that:
\begin{eqnarray}
- \tr\left(S \frac{ {\cal W}^\beta}{z-\Phi}A \frac{ \Phi^T{\cal W}_\beta^T}{z-\Phi^T}\right) &=&
\tr\left(S \frac{\Phi {\cal W}^\beta}{z-\Phi}A \frac{ {\cal W}_\beta^T}{z-\Phi^T}\right) = \nonumber\\
\tr\left(S \frac{ {\cal W}^\beta}{z-\Phi}A \frac{(z-\Phi^T) {\cal W}_\beta^T}{z-\Phi^T}\right) &\equiv&
-\tr\left(S \frac{ {\cal W}^2}{z-\Phi}A\right)
\end{eqnarray}
where in the last line we used the chiral ring relation $A{\cal
  W}_\beta^T \equiv -{\cal W}_\beta A$.  Combining this with the
anomaly term (\ref{AXY.two.rhs}) leads to the Konishi relation
(\ref{konishi.A.two}).

Let us now turn to transformations of the fundamental fields. Starting
with:
\begin{equation}
\delta Q_f = \sum_{g=1}^{{\hat N}_F} \lambda_{fg} \frac{1}{z-\Phi} Q_g~~,
\end{equation}
we find:
\begin{equation}
2\sum_{g=1}^{{\hat N}_F} \lambda_{fg} Q_f^T S \frac{1}{z-\Phi} Q_g \equiv 
{\bf R}(z) \lambda_{ff}~~
\end{equation}
where we do {\em not} sum over the flavor index $f$. This is the most
general form of the Konishi anomaly without any restriction on
$\lambda_{fg}$.  Since $\lambda_{fg}$ is arbitrary, we conclude:
\begin{equation}
2Q_f^T S \frac{1}{z-\Phi} Q_g \equiv 
{\bf R}(z) \delta_{fg}~~.
\end{equation}
Taking the trace of this equation leads to the Konishi relation
(\ref{konishi.Q.one}).

\end{document}